\def\wgt{\mathop{\rm wgt}}

\def\rank{\mathop{\rm rank}}

\def\css{\mathop{\rm CSS}}

\def\llr{\mathop{\rm LLR}\nolimits}

\documentclass[aps,pra,tightenlines,reprint,
notitlepage,longbibliography]{revtex4-2}

\usepackage{longtable}
\usepackage[T1]{fontenc}
\usepackage[colorlinks,citecolor=blue,driverfallback=pdftex]{hyperref}
\usepackage{bbold}
\usepackage{amsmath}
\usepackage{amssymb}
\usepackage{xcolor}
\usepackage{braket}

\usepackage{amsfonts}
\usepackage{graphicx}
\graphicspath{{./}{./figs/}{./pak/}{../two-shot/figs/}}
\usepackage{tikz}

\usepackage[inline]{enumitem}
\setlist{nosep}

\usepackage{amsthm}

\def\bs#1{\boldsymbol{#1}}

\definecolor{darkpink}{RGB}{219, 112, 147} 

\begin{document}

\title{Single-shot and two-shot decoding with generalized bicycle codes}

\author{Hsiang-Ku Lin}

\affiliation{Department of Physics \& Astronomy, University of
  California, Riveside, California 92521 USA}

\author{Xingrui Liu}

 \affiliation{Department of Physics \& Astronomy, University of
   California, Riveside, California 92521 USA}

\author{Pak Kau Lim}

\affiliation{Department of Physics \& Astronomy, University of
  California, Riveside, California 92521 USA}

 \author{Leonid P. Pryadko}
 \email{leonid.pryadko@ucr.edu}
 \affiliation{Department of Physics \& Astronomy, University of
   California, Riveside, California 92521 USA}
 \date{\today}
 \begin{abstract}
   Generalized-bicycle (GB) and more general two-block group-algebra
   (2BGA) quantum error-correcting codes have naturally redundant
   minimum-weight stabilizer generators.  To use this redundancy, we
   constructed a large number of ``planar'' 2BGA codes over abelian
   groups with one and two generators, with each block row of weight
   3, relatively large dimensions, distances, and maximum syndrome
   distance $d_{\rm S}=3$.  We simulated the performance of three such
   codes under phenomenological noise and standard circuit noise,
   using sliding window sequential decoding protocol covering $T\ge 1$
   measurement rounds at a time, based on an in-house binary BP+OSD
   decoder.  While true single-shot decoding ($T=1$) suffers from a
   significant loss of accuracy, already two-shot ($T=2$) decoding
   gives nearly the same logical error rates as multi-shot with much
   larger $T$.  Comparison with the same codes but additional
   stabilizer generators dropped shows that redundancy significantly
   improves decoding accuracy for all $T\ge1$.
\end{abstract}
\maketitle

\section{Introduction.}

Quantum error correction (QEC) is so far the only known way to
scalable quantum computation.  For calculations big enough to be
useful, coherence must be maintained for millions or even billions of
error-correction cycles.  Every cycle a quantum syndrome extraction
circuit is executed; the resulting syndrome data needs to be processed
by a classical computer.  This classical syndrome-based decoding
problem can be solved sequentially, using measured syndrome data from
a few cycles at a
time\cite{Dennis-Kitaev-Landahl-Preskill-2002,Gong-Cammerer-Renes-2024}.
While off-line decoding is sufficient for Clifford circuits,
non-Clifford gates required for universal quantum computation require
real-time decoding.  Even though some delay can be tolerated at the
cost of increased circuit complexity, shorter lag times are strongly
preferable\cite{Skoric-Browne-Barnes-Gillespie-Campbell-2023}.

From this viewpoint, single-shot
decoding\cite{Bombin-2015,Brown-Nickerson-Browne-2016,Campbell-2018},
where syndrome data from each cycle is processed right after it
becomes available, is strongly preferable.  Indeed, here the
unavoidable physical delay is minimal, and a smaller amount of
syndrome data can give faster decoding.

However, single-shot fault-tolerant QEC requires measuring some
redundant stabilizer generators.  This redundancy helps to control
measurement errors and thus enables independent decoding of each batch
of syndrome data.  Such ideas go back to the seminal
work\cite{Dennis-Kitaev-Landahl-Preskill-2002} of Dennis et al.;
various coding theory aspects have been studied in, e.g.,
Refs.~\onlinecite{Fujiwara-2014,Ashikhmin-Lai-Brun-2014,%
  Ashikhmin-Lai-Brun-2016}.  Numerically, variants of single-shot
decoders with phenomenological error model have been studied for 3D
homological-product
codes\cite{Quintavalle-Vasmer-Roffe-Campbell-2021}, higher-dimensional
hyperbolic codes\cite{Breuckmann-Londe-2020}, higher-dimensional
hypergraph-product (HP) codes\cite{Higgott-Breuckmann-2023}, and also
for regular 2D toric codes with added redundant higher-weight
stabilizer generators\cite{Lin-Huang-Brown-2023}.  In the latter
paper\cite{Lin-Huang-Brown-2023}, also a circuit-based Pauli error
model has been studied but did not show a threshold or much of an
improvement over conventional circuits measuring only minimum-weight
stabilizer generators.

Redundancy of minimum-weight stabilizer generators is a natural
feature of
generalized-bicycle\cite{Kovalev-Pryadko-Hyperbicycle-2013,%
  Panteleev-Kalachev-2019,Wang-Pryadko-2022} (GB) and, more generally,
two-block codes, including abelian and non-abelian two-block group
algebra (2BGA) codes\cite{Wang-Lin-Pryadko-2023,Lin-Pryadko-2023}.
While codes in this family have distances scaling as a square root of
the block length (exact distance scaling is only known for those
permutation-equivalent to HP codes), short 2BGA codes with small
stabilizer generator weights have excellent parameters and are
relatively easy to construct by exhaustive search.

Recently, Bravyi et al.\ have constructed and studied several planar
abelian 2BGA codes\cite{Bravyi-etal-Yoder-2023}.  Their {\em
  bivariate-bicycle\/} (BB) codes, constructed from a pair of
weight-three polynomials with two variables, admit a qubit layout for
a feasible hardware implementation with two layers of superconducting
qubits.  Bravyi et al. constructed near time-optimal fault-tolerant (FT)
syndrome measurement circuits for these codes; simulations using
belief propagation and ordered-statistics decoding (BP+OSD)
demonstrated a pseudothreshold of 0.8\%, just a bit below $\sim 1\%$
threshold values known for the surface
codes\cite{Wang-Fowler-Hollenberg-2011}.  However, the effect of
redundancy of minimum-weight stabilizer generators on decoding
accuracy has not been studied for 2BGA codes. In particular, BB codes in
Ref.~\onlinecite{Bravyi-etal-Yoder-2023} have not been optimized for
syndrome distance: all except the code $[[72,12,6]]$ have syndrome
distances $d_{\rm S}=2$, same as for toric codes.

In this work we study how the natural redundancy of minimum-weight
stabilizer generators of GB codes can be used to improve the accuracy
and speed of syndrome-based decoding in an FT setting.  To this end,
we design short GB codes with row weights $6$ (two weight-three
polynomials, same weights as for codes in
Ref.~\onlinecite{Bravyi-etal-Yoder-2023}) and syndrome code distances
$d_{\rm S}=3$ (largest possible for this weight), construct optimal
measurement circuits for these codes, and simulate these codes in an
FT setting, i.e., with syndrome measurement errors, using both the
phenomenological error model and the circuit error model.  We compare
the accuracy of several sequential decoding schemes, including one-
and two-step single-shot
decoding\cite{Bombin-2015,Campbell-2018,Roffe-White-Burton-Campbell-2020,%
  Quintavalle-Vasmer-Roffe-Campbell-2021,Higgott-Breuckmann-2023} and
more general sliding window (SW) decoding
schemes\cite{Tan-Zhang-Chao-Shi-Chen-2023,%
  Skoric-Browne-Barnes-Gillespie-Campbell-2023,Gong-Cammerer-Renes-2024}.
In particular, we study the decoding accuracy of GB codes with some or
all redundant rows removed, which reduces the syndrome distance.  Even
though the constructed codes have syndrome distances $d_{\rm S}=3$ and
cannot have good
confinement\cite{Quintavalle-Vasmer-Roffe-Campbell-2021}, our results
show that row redundancy improves the accuracy of SW decoding at all
window sizes $T$, and accelerates the convergence with increasing $T$.
In particular, two- or three-shot SW decoding (with $T=2$ and $3$,
respectively) is sufficient to reach near-optimal decoding accuracy.

\section{Methods}
\subsection{Generalized-bicycle codes}
In this work we focus on regular (also single-variate, or
quasi-cyclic), and bi-variate, (or quasi-abelian) GB codes, or,
equivalently, two-block group algebra (2BGA) codes over abelian groups
with one and two generators, see
Refs.~\onlinecite{Kalachev-Panteleev-2020,Lin-Pryadko-2023} for the
details.  These are qubit CSS codes constructed from a pair of binary
commuting square matrices, $AB=BA$, with the stabilizer generator
matrices in the form\cite{Kovalev-Pryadko-Hyperbicycle-2013}
\begin{equation}%
  H_X=\Bigl(A,B\Bigr),\; H_Z^T={B\choose-A}.
\label{eq:css-two-block-matrices}%
\end{equation}%
Explicitly, for quasi-cyclic GB codes, $A\equiv a(P_\ell)$ and
$B\equiv b(P_\ell)$ are circulant matrices, where $P_\ell$ is an
$\ell\times \ell$ cyclic permutation matrix, generator of the
order-$\ell$
cyclic group $C_\ell$, and $a(x),b(x)\in\mathbb{F}_2[x]$
are polynomials with binary coefficients and degrees smaller than
$\ell$.  Similarly, for BB codes over a group
$C_{\ell_X}\times C_{\ell_Y}$ of order $\ell=\ell_X\ell_Y$, the
matrices are constructed from a pair of two-variable polynomials,
$a(x,y), b(x,y)\in \mathbb{F}_2[x,y]$,
$$
A=a(P_X,P_Y),\quad B=b(P_X,P_Y),
$$
with commuting permutation matrices
$P_X\equiv P_{\ell_X}\otimes I_{\ell_Y}$ and
$P_Y\equiv I_{\ell_X}\otimes P_{\ell_Y}$, where $I_\ell$ is the
$\ell\times \ell$ identity matrix and `$\otimes$' denotes the
Kronecker product.  Evidently, in all cases, the block length of the
CSS code (\ref{eq:css-two-block-matrices}) is $n=2\ell$.  Moreover,
for quasi-abelian GB codes, the CSS
matrices~(\ref{eq:css-two-block-matrices}) have equal ranks,
\begin{equation}
  \label{eq:css-ranks}
  \rank H_X=\rank H_Z\equiv\ell-\kappa,
\end{equation}
and the code dimension $k=2\kappa$ is
even\cite{Kalachev-Panteleev-2020,Lin-Pryadko-2023}.  In addition, the
two CSS distances of quasi-abelian GB codes coincide, $d_X=d_Z$.

Quasi-cyclic GB codes are easier to analyze algebraically.  In
particular, the code dimension $k$ is
given\cite{Panteleev-Kalachev-2019} in terms of the auxiliary
polynomial
\begin{equation}
h(x)=\gcd(a,b,x^\ell-1),\;\text{namely,}\;\; k=2\deg h(x).\label{eq:h-poly}
\end{equation}
While there is no simple expression for the distance, an upper bound
is given in terms of the distance of the cyclic code with the check
polynomial (\ref{eq:h-poly}), $d\le d_h^\perp\equiv d(C_h^\perp)$.
Additional distance bounds can be found in
Ref.~\onlinecite{Wang-Pryadko-2022}.

\subsection{Syndrome code}
\label{sec:syndrome-code}

The CSS generator matrices $H_X$ and $H_Z$ in
Eq.~(\ref{eq:css-two-block-matrices})
 automatically satisfy the row orthogonality conditions,
\begin{equation}
  \label{eq:CSS-orthogonality}
  H_X H_Z^T=0;
\end{equation}
both matrices have $\ell$ rows and ranks $\ell-\kappa$, i.e., each has $\kappa$
redundant rows.  Thus, e.g., valid $X$-syndrome vectors, obtained by
measuring the stabilizer generators defined by rows of $H_X$, form a
binary linear code, the $X$-syndrome code, of length $\ell$, dimension
$\ell-\kappa$, and distance $d_{\rm S}$.  In general, due to symmetry, for
any 2BGA code, $d_{\rm S}\ge2$.  On the other hand, the distance
$d_{\rm S}$ of a syndrome code cannot exceed the minimum column weight
of the corresponding CSS matrix,
\begin{equation}
  d_{\rm S}\le \min\left(w_a,w_b\right),\;\, w_a\equiv \wgt a, \;\,
  w_b\equiv \wgt b,
  \label{eq:bound-ds}
\end{equation}
where the polynomial weight $\wgt a$ is the number of non-zero
coefficients.

For any single-variate GB code, the syndrome code is a cyclic code
generated by the polynomial $h(x)$ in Eq.~(\ref{eq:h-poly}).  The
corresponding check matrix $M_X$ (and similarly defined check matrix
$M_Z$ for the $Z$-syndrome code) can be constructed from the
polynomial $g(x)\equiv(x^\ell-1)/h(x)$,
\begin{equation}
  M_X=M_Z^T=g(P_\ell),\quad M_X H_X=0,\quad H_Z^TM_Z^T=0.
  \label{eq:metacheck-matrices}
\end{equation}
Rows of $M_X$ and $M_Z$, respectively, correspond to $X$ and $Z$
metachecks\cite{Campbell-2018}.  The syndrome code distance
$d_{\rm S}$ is different from the single-shot distance $d_{\rm SS}$
defined in Ref.~\onlinecite{Campbell-2018} which in our case is
infinite since the syndrome code contain only valid syndrome vectors.
Moreover, for GB codes of interest (i.e., $k$ and $d$ large, with
bounded $w_a$ and $w_b$), in general no bounded-weight metachecks can
be found.  In particular, in the case of single-variate GB codes, from
general theory of cyclic codes (see, e.g., Ref.~\onlinecite{MS-book})
it follows that when $d_{\rm S}$ is small compared to $\ell$, the
corresponding dual code likely has a large distance.  That is, for any
low density parity-check (LDPC) GB code constructed from a pair of
circulant matrices, the corresponding $X$ and $Z$ syndrome codes are
cyclic codes generally without the LDPC property.  This is different
from, e.g., higher dimensional topological
codes\cite{kitaev-anyons,Bravyi-Kitaev-1998,Freedman-Meyer-1998,%
  Dennis-Kitaev-Landahl-Preskill-2002,Bombin-MartinDelgado-2007,%
  Castelnovo-Chamon-2008,Mazac-Hamma-2012,%
  Bombin-Chhajlany-Horodecki-MartinDelgado-2013} or higher-dimensional
HP codes\cite{Tillich-Zemor-2009,Campbell-2018,Zeng-Pryadko-2018,%
  Zeng-Pryadko-hprod-2020}.

\subsection{Error models}
\label{sec:errors}

A {\bf phenomenological noise model} is an idealized and heavily
simplified model which includes both data-qubit and measurement
errors, where the former errors accumulate with time while the latter
errors happen independently and have no effect on the data qubits.

For CSS codes of interest here, we consider the simplest case of
$N$-times repeated measurements of $X$-generators with independent
identically distributed (iid) $Z$ errors $e_{ij}$ (on qubit $i$ and
measurement round $j\le N$) with probability $p$, and iid measurement
errors $\epsilon_{ij}$ (on syndrome bit $i$ in measurement round
$j<N$) with probability $q$.  It is assumed that the last round has no
measurement error, in agreement with an experimental protocol where
data qubits are measured in the last round in the $X$ basis: these
measurement results can be used to calculate the syndrome
without an error.

Given a qubit CSS code $[[n,k,d]]$ with binary matrices $H_X$ (size
$r \times n$) and $H_Z$ (size $r'\times n$) such that $H_X H^T_Z=0$,
we introduce binary data $\bs{e}_j\in \mathbb{F}_2^n$ and syndrome
$\bs{\epsilon}_j\in \mathbb{F}_2^{r}$ error vectors that occurred in
the $j$\,th round of measurements (these vectors are formed by the
phase-error bits $e_{ij} $ and measurement-error bits $\epsilon_{ij}$,
respectively).  Since data qubit errors accumulate, the syndrome
measured in the $j$\,th measurement round is
$$
\bs {s}_j =(\bs e_1+\bs e_2+\ldots+\bs e_j)H_X^T+\bs\epsilon_j, \quad
\bs \epsilon_N\equiv\bs 0,
$$
and the errors can be detected locally by the differences%
$$\bs \sigma_j\equiv \bs {s}_j-\bs {s}_{j-1} = \bs e_j H_X^T+\bs
\epsilon_j-\bs\epsilon_{j-1},\quad \bs \epsilon_0\equiv\bs 0.
$$

It is convenient to combine data errors and measurement errors in two
separate blocks,
\begin{equation}
  {\bf e}=\bigl(\bs e_1,\bs e_2,\ldots,\bs e_N\big|\bs\epsilon_1,\bs
  \epsilon_2,\ldots,\bs \epsilon_{N-1}\big.\bigr)\in\mathbb{F}_2^{Nn +(N-1)r}.
  \label{eq:total-error}
\end{equation}
The combined vector
${\bs\sigma}=(\bs \sigma_1,\bs \sigma_2,\ldots, \bs \sigma_N)$ of
triggered detectors can be computed using the check matrix
\begin{equation}
  \label{eq:HP-Hx}
  {\sf H}_X=\bigl(I_N\times H_X \big| R_N^T\times I_r\big.\bigr),
  \quad \bs \sigma={\bf e}\,  {\sf H}_X^T.
\end{equation}
Two blocks of the matrix ${\sf H}_X$ are written as Kronecker
products, where $R_N$ is a check matrix of the classical repetition
code with distance $N$; it has the dimensions $(N-1)\times N$ and the
elements $[R_N]_{ij}=\delta_{i,j}-\delta_{i,j-1}$.
Eq.~(\ref{eq:HP-Hx}) has the structure of matrices encountered in HP
codes\cite{Tillich-Zemor-2009,Zeng-Pryadko-2018,Zeng-Pryadko-hprod-2020}.
The corresponding degeneracy-generating matrix (for details, proofs,
and a presentation using chain complex notations see
Ref.~\onlinecite{Zeng-Pryadko-hprod-2020}) can be written as
\begin{equation}
  \label{eq:HP-Hz}
  {\sf H}_Z=\left(
    \begin{array}{c|c}
      R_N\times I_n&-I_{N-1}\times H_X^T\\ \hline
      I_{N}\times H_Z& 0
    \end{array}
  \right),
\end{equation}
which immediately gives ${\sf H}_X {\sf H}_Z^T=0$.  Indeed, each row
in the first row block of Eq.~(\ref{eq:HP-Hz}) corresponds to a
trivial undetectable error starting as a data qubit error, followed by
measurement errors in each adjacent generator making it invisible,
followed by an identical data qubit error in the next measurement
round fixing the original error.  Errors in the second row block
represent the degeneracy of the original code: the corresponding rows
are phase errors in the pattern of $Z$ stabilizer generators of the
original code, in each measurement round.  Altogether, the ``big''
quantum code with CSS stabilizer generator matrices (\ref{eq:HP-Hx})
and (\ref{eq:HP-Hz}) has both the dimension $k$ and the distance $d_Z$
of the original quantum
code\cite{Zeng-Pryadko-2018,Zeng-Pryadko-hprod-2020}.  For
completeness, if $L_X$ and $L_Z$ are full-row-rank $k\times n$
matrices generating the logical operators of the original CSS code,
the corresponding matrices of the big code can be written as
\begin{eqnarray}
  \label{eq:HP-Lx}
  {\sf L}_X&=&\bigl(\bs 1_N\times L_X\big|0_{k\times (N-1)r}\big.\bigr),\\
  {\sf L}_Z&=&\bigl(L_Z,0_{k\times (N-1)n}\big|0_{k\times (N-1)r}\big.\bigr).
  \label{eq:HP-Lz}
\end{eqnarray}
Here $0_{a\times b}$ is an $a\times b$ zero matrix,
$\bs1_N\equiv (1,1,\ldots ,1)$ is the unique non-zero codeword of the
repetition code, $\bs 1_N R_N^T=\bs 0$, and the original matrices
$L_X$ and $L_Z$ must have $k$ rows and satisfy the orthogonality conditions,
\begin{equation}
L_XH^T_Z=0,\quad L_ZH^T_X=0,\quad \rank (L_XL^T_Z)=k.\label{eq:CSS-L}
\end{equation}
The corresponding conditions for the big code matrices
(\ref{eq:HP-Lx}) and (\ref{eq:HP-Lz}) are easy to verify directly.

The {\bf circuit error model} we use is the ``standard'' circuit-level
depolarizing noise model nearly identical to those used for surface
code simulations, e.g., in
Refs.~\onlinecite{Raussendorf-Harrington-2007,Wang-Fowler-Hollenberg-2011,%
  Huang-Newman-Brown-2020}.  Namely, given an ideal quantum circuit
formed by 1- and 2-qubit Clifford gates, and single-qubit reset \&
measurement operations, we insert iid single-qubit Pauli $X$ error
after each reset and before each measurement, as well as 1- and
2-qubit depolarizing errors after each 1- and 2-qubit gate,
respectively, including trivial gates (idling).  The idling errors
during each ancillary qubit reset \& measurement include those for the
data qubits at the start of each measurement round.  We note that
except for reset and measurement, our circuits do not have any
single-qubit gates as they are constructed using $X$- and
$Z$-controlled $X$ gates, where the latter is the usual {\tt CNOT}
gate.

Constructed measurement circuits were analyzed and simulated using the
{\tt Stim} package\cite{Gidney-2021-stim}.  For each code, a one-round
measurement circuit was designed, see below in
Sec.~\ref{sec:measurement-circuit}.  Then, two variants of each
simulation were constructed, with data qubits initially prepared in
$\ket0$ or $\ket+$ states, and the final data-qubit measurement done
in the $Z$ or $X$ basis, respectively.  Each circuit included initial
data qubit preparation, $N-1$ identical measurement rounds, followed
by a final round of data qubit measurement.  The complete {\tt Stim}
simulation also includes annotations for the \emph{detector events}
representing relations between measurement in an ideal error-free
case, and the \emph{observables} which are used to check whether the
decoding was correct.  For example, detector events could be the
differences between measurement results in the subsequent rounds of
stabilizer measurements, or similar relations between the $X$ or $Z$
stabilizer measurements in the last round and the final data qubit
measurement results.  Similarly, the observables correspond to the
rows of the logical operator matrices $L_X$ or $L_Z$ such that the
expected values of the corresponding operators in the absence of
errors can be predicted from the qubit initialization and the final
measurement results.

The detector error model (DEM) which can be exported by {\tt Stim} is
a list of all error classes flipping different sets of detector events
or observables, along with their cumulative probabilities $p_i$.
Probabilities of different errors in the same class (producing
identical outcomes) are always combined assuming the corresponding
events independent.  Respectively, the overall error, a set of error
classes that have occurred in a particular simulation, can be
described by a binary vector $\bf e$ whose components $e_i$
have independent Bernoulli distributions with probabilities $p_i$.  If
we combine the detector events and the observables into a parity-check
matrix $\sf H$ and the logical operator matrix $\sf L$, the
corresponding vectors can be written as
${\bs\sigma}={\bf e}\,{\sf H}^T$ and ${\bs\tau}={\bf e}\,{\sf L}^T$.

As a special very simple example, the phenomenological error model
with matrices $\sf H=\sf H_X$ in Eq.~(\ref{eq:HP-Hx}) and
$\sf L=\sf L_X$ in Eq.~(\ref{eq:HP-Lx}), up to a permutation of
columns, is obtained in the presence of (i) phase errors on data
qubits at the start of each measurement round and (ii) phase-flip
errors right before each $X$-basis measurement.  Not surprisingly,
measured observables (rows of matrix $\sf L_X$) have parities defined
by the net accumulated data-qubit $Z$-errors,
$\bs e_{\rm tot}=\bs e_1+\bs e_2+\ldots+ \bs e_N$.

We note that rows of matrix $\sf L$ should be linearly independent
from those of $\sf H$, but the total rank,
$\rank {\sf H}+\rank {\sf L}$, is usually smaller than the number of
columns in these matrices.  Respectively, there are non-zero vectors
orthogonal to rows of both matrices, i.e., trivial undetectable errors
that have no effect on the observables.  These are related to the
circuit error-equivalence group (EEG) generated by error combinations
that have no effect, e.g., a Pauli $X$ on a control qubit just before
a {\tt CNOT} gate, followed (just after the gate) by Pauli $X$ errors
on both qubits in the support of the gate\cite{Pryadko-2020}.  The
corresponding generator matrix ${\sf G}={\sf H}_Z$ together with the
original check matrix ${\sf H}={\sf H}_X$, in effect, form a CSS code
associated with the circuit.  This code is closely related but is not
identical to the subsystem code associated with a measurement
circuit\cite{Bacon-Flammia-Harrow-Shi-2017}.

\subsection{Design of measurement circuits}
\label{sec:measurement-circuit}

A big advantage of square-lattice surface codes is that all stabilizer
generators can be measured at the same time, using a (nominally
non-fault-tolerant) measurement circuit with only one ancillary qubit
per generator.  This is achieved by carefully choosing the order of
addressing individual qubits in $X$ and $Z$ generators so that (i)
there is no overlap between qubit pairs addressed at each time step;
(ii) measurement circuit be valid, i.e., no unwanted entanglement be
generated between ancillary qubits; and (iii) so-called hook errors,
weight-two data-qubit errors generated by ancillary-qubit faults in
the middle of the circuit, be aligned in the direction perpendicular
to matching logical
operators\cite{Dennis-Kitaev-Landahl-Preskill-2002}.  Numerically,
fault-tolerance can be verified by computing the distance $d_{\rm C}$
of the quantum CSS code associated with the DEM of the measurement
circuit, see Sec.~\ref{sec:errors}.

Our approach for constructing circuits for GB codes is to generalize
the $\sf Z$-$\sf N$ addressing scheme\cite{Tomita-Svore-2014} for
surface codes, relying on high symmetry of the codes.  We start with
the case of $w_a=2$ and any $w_b\ge2$.  The no-overlap condition (i)
is easily resolved if at the same time step we address the qubits
corresponding to the same monomial in the polynomial $a$ or $b$;
different terms in, e.g., $a$ can be chosen for $X$ and $Z$
generators since the corresponding qubits are located in different
blocks, see Eq.~(\ref{eq:css-two-block-matrices}).  The circuit validity
condition (ii) is guaranteed if we address the two monomials of $a$
only in the first and in the last time steps of the circuit, and use
opposite orders for $X$ and $Z$ generators, e.g., address $a_0$ in the
first interval and $a_1$ in the last for $X$ generators, and v.v.\ for
$Z$ generators [here $a(x)=a_0+a_1$ is written as a sum of two
monomials.]  During the remaining time steps
$\{2, 3, \ldots , 1+w_b\}$, any two permutations can be used for the
order of addressing the monomials of $b$ for $X$ and $Z$ generators,
respectively.  This pattern always gives valid circuits, and in the
special case $w_b=2$ the familiar ${\sf Z}$-${\sf N}$ patterns are
recovered (although not necessarily oriented correctly to control hook
errors.)

With $w_a=w_b=3$, time-optimal circuits do not exist.  However, if we
shift, say, the $Z$-measurement circuits by one time step, it is easy
to get a family of depth-7 circuits similar to those in
Ref.~\onlinecite{Bravyi-etal-Yoder-2023}.  For example, for $X$
generators, address monomials $a_0$ and $a_1$ of $a$ in the 1st and
2nd steps, the remaining monomial $a_2$ in the 6th, and none in the
last, while for $Z$ generators, skip the 1st step, address $a_2$ in
the 2nd step, and $a_0$ and $a_1$ during the 6th and 7th steps, in any
order.  The monomials of $b$ should be addressed during 3rd, 4th,
and 5th steps, using arbitrary but fixed permutations for $X$ and $Z$
generators.

While this prescription guarantees the validity of constructed
measurement circuits, the circuit distance does not necessarily reach
that of the original code.  Just as in the case of surface codes, the
remaining freedom in ordering the terms can be used to optimize the
circuit distance.  For GB codes with $w_a=w_b=3$ the hook errors have
weights up to 3.  It is quite unexpected that with a careful circuit
construction these may act as weight-one errors.  The optimal
measurement schedules, i.e., the time steps when the qubits
corresponding to different monomials of $a$ and $b$ should be
addressed, are given for each code in columns labeled ``addr'' of
Tab.~\ref{tab:codes} , and is also shown in Fig.~\ref{fig:layout} for
the case of $\ell=31$ code.

We note that the described protocol does not always result in
fault-tolerant circuits; on occasion, only circuits with distances
smaller than that of the original code can be found.  In such a case,
our solution is to take a different code, e.g., constructed from an
inequivalent pair of polynomials.

\subsection{Syndrome-based decoding}
\label{sec:decoding}

\noindent{\bf General formulation:} We consider the problem of
syndrome-based decoding as that of binary minimum-energy decoding for
a code $\css({H_X,H_Z})$ with parameters $[[n,k,d_X/d_Z]]$, where the
CSS generator matrices $H\equiv H_X$, $G\equiv H_Z$ and logical
generator matrices $L\equiv L_X$, $L_Z$ satisfy the usual
orthogonality conditions, Eqs.~(\ref{eq:CSS-orthogonality}) and
(\ref{eq:CSS-L}).  Error vectors $\bs e\in \mathbb{F}_2^n$ are assumed
to have independent components $e_i\in \mathbb{F}_2$, Bernoulli random
variables with parameters $0<p_i<1$, $i\in\{1,2,\ldots,n\}$;
probability of such a vector is a strictly decreasing function of the
corresponding \emph{energy}
\begin{equation}
  \label{eq:energy}
  E(\bs e)\equiv \sum_{i=1}^n e_i \llr_i ,\quad \llr_i\equiv \ln {1-p_i\over p_i}.
\end{equation}
Namely, given a syndrome vector $\bs s=H^T\bs e$ for some unknown
binary error vector $\bs e\in \mathbb{F}_2^n$, we want to find a
minimum-energy error vector $\hat{\bs e}$ matching the syndrome,
$\bs s=H^T\hat{\bs e}$.  The decoding is considered successful if and
only if the observables vectors coincide, $\hat{\bs\tau} =\bs \tau$, with
\begin{equation}
  \bs\tau\equiv L^T {\bs e},\quad \hat{\bs\tau} \equiv L^T\hat{\bs e}.\label{eq:obs-match}
\end{equation}

Note that, ideally, optimal decoding for a quantum code with
$\rank G\neq 0$, corresponds to finding the most likely \emph{error
  degeneracy class} which contains mutually degenerate error vectors;
this problem can be formulated as that of \emph{minimal free energy}
(MFE)
decoding\cite{Dennis-Kitaev-Landahl-Preskill-2002,Kovalev-Pryadko-SG-2015}.
We use suboptimal minimum-energy decoding because of the much higher
complexity of the MFE decoding\cite{Iyer-Poulin-2013}.  However, in
the practically important region of small error rates we do not expect
much of an advantage from MFE decoding.

We should also note that a variety of related decoding problems with
correlated errors can be formulated as binary decoding with
independent errors, in particular, decoding with general one- and
few-qubit Pauli errors, as well as various fault-tolerant decoding
problems, e.g., with phenomenological or circuit-based errors, see
Sec.~\ref{sec:errors}.  However, maps between different
codes\cite{Pryadko-2020} which rely on the properties of multi-variate
Bernoulli distributions are only exact for free-energy-based decoding.
More over, with iterative decoders like belief propagation (BP), the
convergence may vary dramatically for nominally-equivalent codes.
Same is true even when the minimization problems are equivalent, e.g.,
when the decoding problem can be rewritten in terms of higher
alphabets, whether due to the structure of the
code\cite{Andriyanova-Maurice-Tillich-2012,Xie-Yuan-2016}, or when the
error model has additional symmetry, as in the case of quaternary BP
with depolarizing noise\cite{Kuo-Lai-2020}.

\noindent{\bf BP decoder:} Iterative decoders, including min-sum and
belief-propagation decoders are a de-facto industry
standard for classical LDPC codes used for communications, as they
achieve linear complexity and may have near zero error rates close to
the Shannon limit\cite{Richardson-Shokrollahi-Amin-Urbanke-2001,%
  Chung-Forney-Richardson-Urbanke-2001}.

In the case of degenerate quantum LDPC codes, the accuracy of these
decoding algorithms is plagued by convergence
problems\cite{Poulin-Chung-2008}.  Namely, the decoding error
probabilities may be very low if the algorithm converges, but the
convergence failures are common, and their likelihood may actually
increase with increasing code distances, driving up the overall error
rates.  Panteleev and Kalachev\cite{Panteleev-Kalachev-2019} suggested
a partial resolution, the ordered-statistics
decoder\cite{Fossorier-Lin-1995,Fossorier-2001} (OSD) used for
post-processing.  As the input, OSD uses the reliability data
resulting from prior failed BP decoding attempt, and has a complexity
scaling as $\mathcal{O}(n^3)$.  For a degenerate quantum LDPC code
with stabilizer generators of minimum weight $w\le d$, BP fail rates
typically scale as $n p^{t_{\rm cl}}$, where $w$ is distance of the
corresponding classical code and $t_{\rm cl}=\lceil w/2\rceil$ is the
number of errors it can correct.  Thus, at small $p$ and $w\ge 6$, the
OSD decoder is triggered not frequently enough to substantially
degrade the average decoding speed\cite{Panteleev-Kalachev-2019}.

Unfortunately, in the case of circuit error models, each two-qubit
gate is associated with two or more distinct trivial errors of weight
$w=3$ (these generate the EEG of the circuit, see
Sec.~\ref{sec:errors}).  Respectively, the total number $n$ of
distinct fault patterns (variable nodes associated with the DEM) is
large; it scales with the area of the measurement circuit.  Thus, BP
convergence-fail rates $\mathcal{O}(np^2)$ are not particularly small
for $p$ values of interest, which substantially reduces the speed of
the combined BP+OSD decoding for circuit error models.

To accelerate the decoding, we added a pre-decoding step, which tries
to match each non-zero syndrome vector as a collection of isolated
small-weight errors.  To this end, non-zero syndrome bits are grouped
into clusters according to the connectivity graph associated with the
check matrix (neighboring check nodes are adjacent to the same
variable node).  Syndrome vector of each cluster is then searched in a
hash with pre-computed collection of error-syndrome pairs.
Pre-decoding is considered successful only if all syndrome clusters
can be matched; otherwise regular BP+OSD decoding is done.  For
smaller error rates, over 90\% of all syndrome vectors can be decoded
this way, substantially increasing the average decoding speed.

The described algorithm is implemented as a part of the software
package {\tt vecdec}\cite{Pryadko-2025-vecdec} available at GitHub.
In addition to cluster-based pre-decoder and several variants of BP
decoder with OSD post-processing, the package also implements a
minimum-weight decoder using random information set (RIS) algorithm
similar to those used for classical
codes\cite{Prange-1962,Leon-1988,Kruk-1989,Coffey-Goodman-1990}; see
Ref.~\onlinecite{Dumer-Kovalev-Pryadko-IEEE-2017} for a related
discussion in application to quantum LDPC codes.

For all simulations in this work, we used {\tt vecdec} as the decoding
engine, with pre-decoder attempting to match the syndrome as a
collection of independent weight-one errors, $V$-based serial BP
schedule, using both instantaneous LLRs and LLR values averaged over
several BP steps, and level-1 OSD for post-processing.  Some of the
results were also verified against several alternative decoders,
including BP+OSD package by Joschka
Roffe\cite{Roffe-White-Burton-Campbell-2020,Roffe_LDPC_Python_tools_2022},
{\tt vecdec} in the RIS mode, and yet another BP+OSD decoder written
by XL.

\noindent{\bf Sequential decoding:}
As discussed in Sec.~\ref{sec:errors}, in all simulated quantum circuits
data qubits are measured  at the end, which allows to verify
whether the decoding was successful.  The corresponding detector error
models, whether constructed by hand or exported from {\tt Stim}, can be seen
as asymmetric quantum CSS codes.  Thus, syndrome data from
each simulation run can be decoded in a single step.  While this
method does not scale well for simulations or experiments with many
measurement rounds, it gives the most accurate decoding.

An alternative is a sequential one-step decoding
pro\-to\-col\cite{Breuckmann-Londe-2020,%
  Grospellier-Groues-Krishna-Leverrier-2020,Higgott-Breuckmann-2023}
which relies on the locality of detector error models.  Namely, a
fault during a measurement round $t$ may affect syndrome measurement
outcomes starting with this ($t$) or next ($t+1$) round, which may
trigger detectors in clusters up to three rounds long.  Respectively,
the check matrices of the corresponding DEMs have distinct band structure,
somewhat similar to spatially-coupled LDPC codes.

This allows for decoding protocols where only syndrome data from $T$
adjacent measurement rounds is used in each decoding step, with the
decoding window shifted forward one round after each step.  The
variable nodes which may affect the ``past'' check nodes above the
current decoding window are frozen, and are only used to update the
check nodes for the current decoding step.  As an illustrative
example, in Fig.~\ref{fig:sw-decoding} (a) and (b) we show matrices
used for partial decoding with $T=1$ and $T=2$, respectively.  Always,
$T=N$, the number of measurement rounds, corresponds to full-block decoding.

\begin{figure}[htbp]
  \centering
  \begin{eqnarray*}
    ({\rm a})\quad {\sf H}
    &=& \begin{tikzpicture}[baseline={(0, 0)}]
          \node at (0, 0) {$
            \left(
              \begin{array}{ccccccccccc}
                H_X&I& & & & & & & & & \\
                   &I&H_X&I& & & & & & & \\
                   & & &I&H_X&I& & & & & \\
                   & & & & &I&H_X&I& & & \\
                   & & & & & & &I&H_X&I& \\
                   & & & & & & & & &I&H_X\\
              \end{array}
            \right)
            $};
          \draw[rounded corners, thick, gray]
          (-3.1, 1.4) rectangle (-2.0, 0.92);
          \draw[rounded corners, thick, gray]
          (-2.0, .92) rectangle (-0.88, 0.46);
          \draw[rounded corners, thick, gray]
          (-0.88, 0.46) rectangle (0.23, 0.);
          \draw[rounded corners, thick, gray]
          (0.23, 0.) rectangle (1.34, -0.45);
          \draw[rounded corners, thick, gray]
          (1.34, -0.45) rectangle (2.42, -0.9);
          \draw[rounded corners, thick, gray,dashed]
          (2.42, -0.9) rectangle (3.09, -1.35);
        \end{tikzpicture}\\
    ({\rm b})\quad {\sf H}
    &=& \begin{tikzpicture}[baseline={(0, 0)}]
      \node at (0, 0) {$
        \left(
          \begin{array}{ccccccccccc}
            H_X&I& & & & & & & & & \\
               &I&H_X&I& & & & & & & \\
               & & &I&H_X&I& & & & & \\
               & & & & &I&H_X&I& & & \\
               & & & & & & &I&H_X&I& \\
               & & & & & & & & &I&H_X\\
          \end{array}
        \right)
        $};
      \draw[rounded corners, thick, gray]
      (-3.1, 1.4) rectangle (-.9, 0.45);
      \draw[rounded corners, thick, gray]
      (-2.0, .95) rectangle (0.23, -.02);
      \draw[rounded corners, thick, gray]
      (-0.88, 0.5) rectangle (1.34, -0.46);
      \draw[rounded corners, thick, gray]
      (0.23, 0.) rectangle (2.61, -.97);
      \draw[rounded corners, thick, gray, dashed]
      (1.34, -0.45) rectangle (3.09, -1.35);
    \end{tikzpicture}
  \end{eqnarray*}
  \caption{SW decoding with phenomenological error model including
    $N=6$ rounds of measurements: (a) single-shot one-step SW
    decoding, $T=1$; (b) two-shot one-step decoding, $T=2$. Matrix
    ${\sf H}$ is obtained from ${\sf H}_X$ in Eq.~(\ref{eq:HP-Hx}) by
    a column permutation.  Framed boxes (``windows'') indicate the
    check submatrices used in different decoding steps, starting from
    the top.  Variable nodes corresponding to columns to the left of
    the current window are frozen; these are used to update check node
    values for the current decoding step.}
  \label{fig:sw-decoding}
\end{figure}

When redundant stabilizer generators are present, one may also use
single-shot two-step decoding\cite{Campbell-2018}.  Here, at the first
decoding step, syndrome measurement errors are corrected, and at the
second, the minimum-energy error for data qubits is found assuming the
corrected syndrome is error-free.

According to Ref.~\onlinecite{Quintavalle-Vasmer-Roffe-Campbell-2021},
this method works well for code families where syndrome weight grows
with the weight of the error (such a property is characterized as
\emph{confinement}).  This is not the case for the codes in
Tab.~\ref{tab:codes}: some errors of weight 1, 2, and 3 give syndrome
weights 3, 4, and 3, respectively.  That is, linear confinement is
present only for errors of maximum weight two.

\section{Results}
\subsection{Constructed codes}
Given relatively high circuit thresholds for BB codes in
Ref.~\onlinecite{Bravyi-etal-Yoder-2023}, we constructed a number of
single- and bi-variate GB codes from weight-three polynomials,
$w_a=w_b=3$, which gives stabilizer generators of weight $w=6$.  All
such codes admit feasible two-dimensional layouts, and will be later
denoted for short as BB(3,3) codes.  To analyze the effect of syndrome
redundancy, we focused on ``better'' syndrome codes with higher
syndrome distances.  With weight-three polynomials, the only
possibilities are $d_{\rm S}=2$ (majority of the codes), and
$d_{\rm S}=3$.

Given that any group $C_m\times C_n$ with $\gcd(m,n)=1$ is isomorphic
to a cyclic group $C_\ell$ of the same order, $\ell=nm$, we considered
cyclic groups $C_\ell$ of order $\ell$, and two-generator abelian
groups $C_m\times C_n$ of order $\ell=nm$ with $\gcd(m,n)>1$.
Specifically, for all such groups of orders $3\le \ell\le 75$ (and
some of the larger groups), we exhaustively enumerated parameters of
non-trivial ($k\neq0$) BB(3,3) codes with distances $d>2$, and
syndrome distances $d_{\rm S}=3$, using an enumeration algorithm
similar to that in Ref.~\onlinecite{Lin-Pryadko-2023}.

The GAP package {\tt
  QDistRnd}\cite{Pryadko-Shabashov-Kozin-QDistRnd-2021} was used to
calculate initial upper bounds on the distance.  Subsequently, the
code parameters, including distances and syndrome distances have been
verified with the package {\tt vecdec}\cite{Pryadko-2025-vecdec}; for
codes with distances $d\le 10$, the results of the stochastic
algorithm have been verified with the program {\tt
  dist\_m4ri}\cite{Pryadko-Zeng-dist-2024} which implements a
deterministic cluster-based codeword enumeration
algorithm\cite{Dumer-Kovalev-Pryadko-IEEE-2017} suitable for (quantum
or classical) LDPC codes.  The parameters and the generating
polynomials for the designed codes are listed in
Tab.~\ref{tab:big-codes} in the Appendix.  The verification script is
available at the GitHub repository\cite{Lin-Liu-Lim-Pryadko-github-2shot}.

On general grounds, with sufficiently large $\ell$, $k$, and $d$, we
expect many properties of the Tanner graphs corresponding to the CSS
matrices $H_X$ and $H_Z$ constructed from same-weight polynomials to
be similar.  To ensure even more similarities in the sequence of the
codes we studied, we picked three codes constructed with cyclic groups
of order $\ell=2^m-1$, with $m=4,5,6$; the corresponding syndrome
codes are primitive BCH codes.  For convenience, the parameters of
these three codes are listed separately in Tab.~\ref{tab:codes}, along
with some additional information relevant for circuit simulations.

\begin{figure}[htbp]
\includegraphics[width=2.4in]{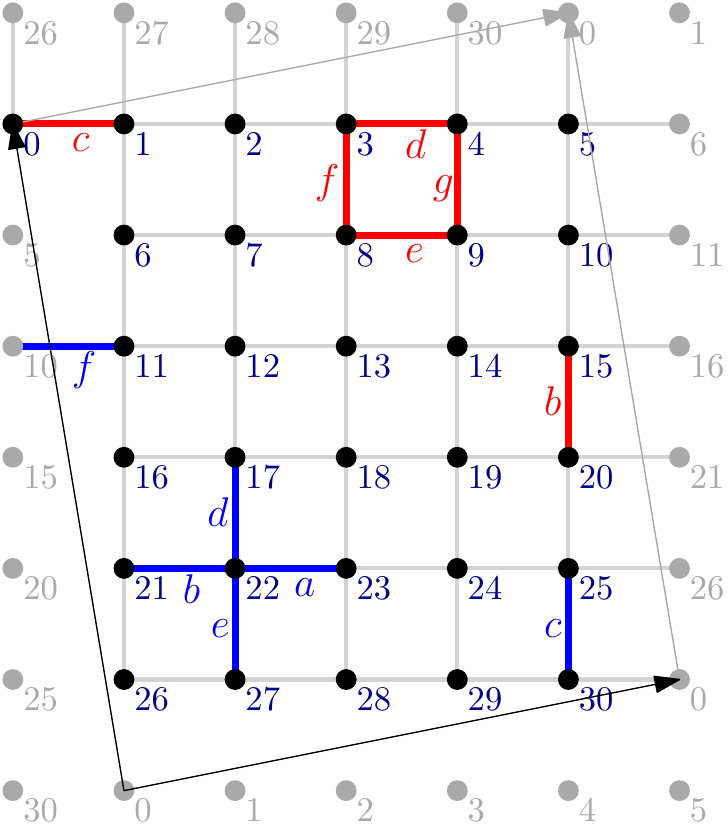}
\caption{(Color online) A planar arrangement of the GB code with
  parameters $[[62,10,6]]$ (parameters shown in bold in
  Tab.~\ref{tab:codes}) and syndrome distance $d_{\rm S}=3$.  This layout
  corresponds to a two-generator presentation of the cyclic group,
  $C_{31} \cong \langle x,y|x^5y^{-1}=x y^6=1\rangle$ and equivalent
  polynomials $a'=a(t)$, $b'=t^3b(t)$.  Here qubits are located on the
  horizontal and vertical edges drawn, with adjacent numbers
  indicating positions in each block, arrows show periodicity vectors,
  while a blue and a red edge patterns, respectively, correspond to an
  $X$ and a $Z$ stabilizer generators.  Other stabilizer generators
  are obtained by translations.  Letters from $a$ to $g$ indicate the
  addressing order in the measurement circuit, see
  Tab.~\ref{tab:codes}.}
\label{fig:layout}
\end{figure}

While the codes in Tab.~\ref{tab:codes} are constructed over cyclic
groups, they can be represented as two-variate (planar) codes using
two-generator group presentations and rewriting the original
polynomials $a(t)$, $b(t)$ in terms of new generators.  For groups of
non-prime orders this can be done by decomposing into direct products,
e.g., $C_{15}=C_3\times C_5$.  Otherwise, redundant generators can be
used, giving non-collinear lattice periodicity vectors, see
Fig.~\ref{fig:layout} for an example and
Ref.~\onlinecite{Wang-Pryadko-2022} for details.

\begin{table}[htbp]
  \centering\small
  \begin{tabular}{c|c|c|c|c|c|c|ccc|c|ccc}
    \hline\hline
    $\ell$ & $n$ & $k$ & $d$ &  ${d}_S$& $d_{\rm C}$  & $a(t)$
    &\multicolumn{3}{|c|}{addr} & $b(t)$
    &\multicolumn{3}{|c}{addr}\\ [0.4ex]
    \hline
    15 & 30& 8 & 4& 3&4 & $1+t^6+t^{13}$ &6&2&1& $1+t+t^4$ &3&5&4\\
           &   &   &  &  &  &               &2&6&7&           &3&4&5 \\ \hline
    \bf 31&\bf 62&\bf 10&\bf 6&\bf 3&\bf 6&$1+t+t^{12}$&1&2&6&$1+t^3+t^8$&3&5&4\\
          &      &      &     &     &     &           &6&7&2&           &3&4&5\\ \hline
    63&126&12&10&3&$10^*$&$1+t^7+t^8$&2&6&1&$1+t^{37}+t^{43}$&5&4&3\\
      &   &  &  & &      &          &7&2&6&           &4&5&3\\
  \end{tabular}
  \caption{Studied GB codes with $w_a=w_b=3$: cyclic group order
    $\ell$, code length $n=2\ell$, code dimension $k$, code distance
    $d$, syndrome code distance $d_{\rm S}$, circuit distance
    $d_{\rm C}$, and the polynomials $a(t)$ and $b(t)$.  The columns
    labeled ``addr'' show the circuit time steps for addressing the
    qubits corresponding to individual monomials when measuring $X$
    and $Z$ generators, respectively, see
    Sec.~\ref{sec:measurement-circuit}.  With these circuits, the
    codes saturate the general bounds $d_{\rm S}\le \min(w_a,w_b)$ and
    $d_{\rm C}\le d$.  For the code with $\ell=63$, the superscript
    ``$*$'' indicates that we could not get sufficient statistics to
    assure the absence of non-trivial undetectable circuit errors of
    weight $w<10$.}
  \label{tab:codes}
\end{table}

\subsection{Redundancy and syndrome distance}
In the range studied, we did not find pairs of BB(3,3)
codes with equal parameters but different syndrome distances.
Therefore, to study the effect of redundancy, we considered the
properties of the same codes with some (or none) redundant rows
removed.  While this has no effect on the quantum code itself (code
distance, degeneracy, the non-trivial codewords, etc., remain
unaffected), removal of redundant rows corresponds to puncturing the
syndrome code.

In general, when a classical code is punctured in a single position,
the distance may remain the same or go down by one\cite{MS-book}.  We chose to
remove rows sequentially, which resulted in syndrome distance
decreasing to the minimum $d''_{\rm S}=1$ after the first two steps,
see Tab.~\ref{tab:redund}.

\begin{table}[htbp]
  \centering
  \begin{tabular}[c]{c|c|c|c|c|c}
    rows removed & $d_{\rm S}$ & confinement profile \\ \hline
    0 & 3 & 3,4,3 \\
    1 & 2 & 2,3,2 \\
    2 & 1 & 1,2,1 \\
    3 & 1 & 1,1,1 \\
    4 & 1 & 1,1,1 \\
  \end{tabular}
  \caption{Effect of redundant row removal on the syndrome codes.  In
    the range of error weights shown, $\wgt(e)\le 3$, these properties
    are the same for all constructed codes due to $d_{\rm S}=3$ and
    $d\ge 4$.  Generally, a BB(3,3) code has even dimension,
    $k=2\kappa$ with $\kappa\ge2$; and each CSS matrix has $\kappa$
    redundant rows which may be removed.}
  \label{tab:redund}
\end{table}

We also analyzed the initial confinement
profiles\cite{Quintavalle-Vasmer-Roffe-Campbell-2021}, the dependence
of the minimum syndrome weight on the irreducible weight of the error,
for $\wgt(e)\le t$.  For any BB(3,3) code with $d\ge4$ and
$d_{\rm S}=3$, the profile always starts with $3,4,3,\ldots$: weight-1
error gives syndrome of weight $3$, two such errors sharing a check
give syndrome of weight $4$, and three single-bit errors can always be
found in either block to give syndrome of weight $3$; this follows
from the identity $(a_0+a_1+a_2)^2=a_0^2+a_1^2+a_2^2$ valid for a
weight-3 binary polynomial with monomials $a_0$, $a_1$, and $a_2$.  As
we remove redundant rows, these syndrome weights monotonously decrease
until reaching a flat profile of weight $1$, see
Tab.~\ref{tab:redund}.  We note in passing that confinement profile
increasing only in the range $\wgt(e)\le t=2$ indicates rather weak
single-shot properties of BB(3,3) codes.

In contrast, e.g., for a polynomial in the form of geometrical
sequence, $a=1+a_1+a_1^2$, a weight-$2$ error, e.g., $e=1+a_1$,
results in a weight-$2$ syndrome, i.e., a lower confinement profile.
Syndrome distance $d_{\rm S}=3$ is a sufficient but not necessary
condition to prevent this.  Namely, there are BB(3,3) codes with
$d_{\rm S}=2$ with the same initial confinement profile as in
Tab.~\ref{tab:redund}.

In addition, we studied the dependence of the distribution of
small-weight irreducible codewords for the ``big'' codes corresponding
to $N$-time syndrome measurement in phenomenological error model, see
Eqs.~(\ref{eq:HP-Hx}) and (\ref{eq:HP-Hz}).  It is such codewords that
define the asymptotic small-$p$ fail rates under minimum-energy
decoding\cite{Dumer-Kovalev-Pryadko-bnd-2015}.  As would be expected
from the construction [or from Eq.~(\ref{eq:HP-Lx})], only the net
data-qubit error matters, and the minimum-weight codewords
(non-trivial errors producing zero syndrome) are those supported in a
single measurement round.  To get an irreducible codeword split
between two or more measurement rounds, one must include some syndrome
(measurement) error, which increases the net weight of the codeword.
This gets easier for codes with smaller syndrome distances (or as more
redundant rows are removed), see Tab.~\ref{tab:cw-weight} for an
example.  Here, with the code of distance $d=4$ and syndrome distance
$d_{\rm S}=3$, when $m$ rows are removed,
$A_w^{({\rm big})}=NA_w^{({\rm orig})}$ for $w< d+d_{\rm S}'$, where
$d_{\rm S}'=d_{\rm S}-m$ for $m<3$ and $1$ otherwise is the actual
syndrome distance.  For larger $w$, the numbers of irreducible
codewords generally increases with increasing $m$.  While relatively
minor effect is expected on the error rates with full-block decoding,
this is an indication that added redundancy is beneficial in an FT
regime.

\begin{table}[htbp]
  \centering
  \begin{tabular}[c]{c|c|c|c|c|c|c}
 code & rows removed &   $A_4$& $A_5$ & $A_6$ & $A_7$ & $A_8$ \\ \hline
  original& any & 45 & 0 &  675& 0 & 4635\\ \hline
  $N=6$
&0   &270&0  &4050&1800 &29160 \\
&1   &270&0  &4410&1800 &38892 \\
&2   &270&60 &4778&3380 &52022 \\
&3   &270&190&5180&7440 &69996 \\
&4   &270&420&5622&15502&96664
  \end{tabular}
  \caption{Count $A_w$ of irreducible $Z$-codewords of weight $w$ in
    the original code $[[30,8,4]]$ (top row in Tab.~\ref{tab:codes}),
    and in the corresponding $N=6$ ``big'' phenomenological-model
    code, depending on the number of redundant rows dropped from the
    $H_X$ matrix.  The codes have the same distance $d=4$; $A_w=0$ for
    $w<4$.}
\label{tab:cw-weight}
\end{table}

\subsection{Phenomenological error model results}

We have simulated phenomenological error model with constructed GB
codes, comparing the accuracy of different decoding protocols (See
Sec.~\ref{sec:decoding} for the details of BP decoder implementation
and sequential decoding protocols.)  In the simulations we used the
same data and syndrome error probability $p$ and (unless stated
otherwise) chose the number of repeated measurements equal to the code
distance, $N=d$.

\begin{figure}[htbp] \centering
\begin{minipage}[t]{.95\linewidth} \raisebox{45.mm}[0pt][0pt]{(a)}%
\includegraphics[width=1.\textwidth]{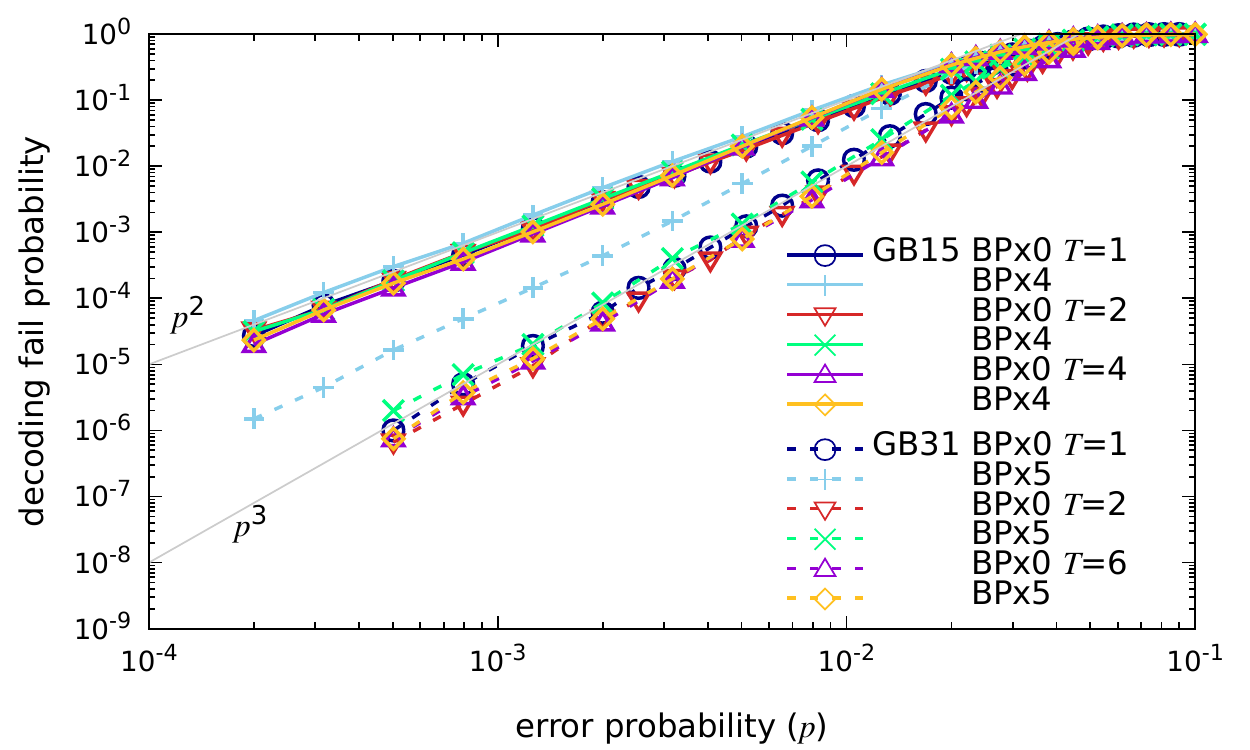}
\end{minipage}
\begin{minipage}[t]{.95\linewidth} \raisebox{45.mm}[0pt][0pt]{(b)}%
  \includegraphics[width=1.\textwidth]{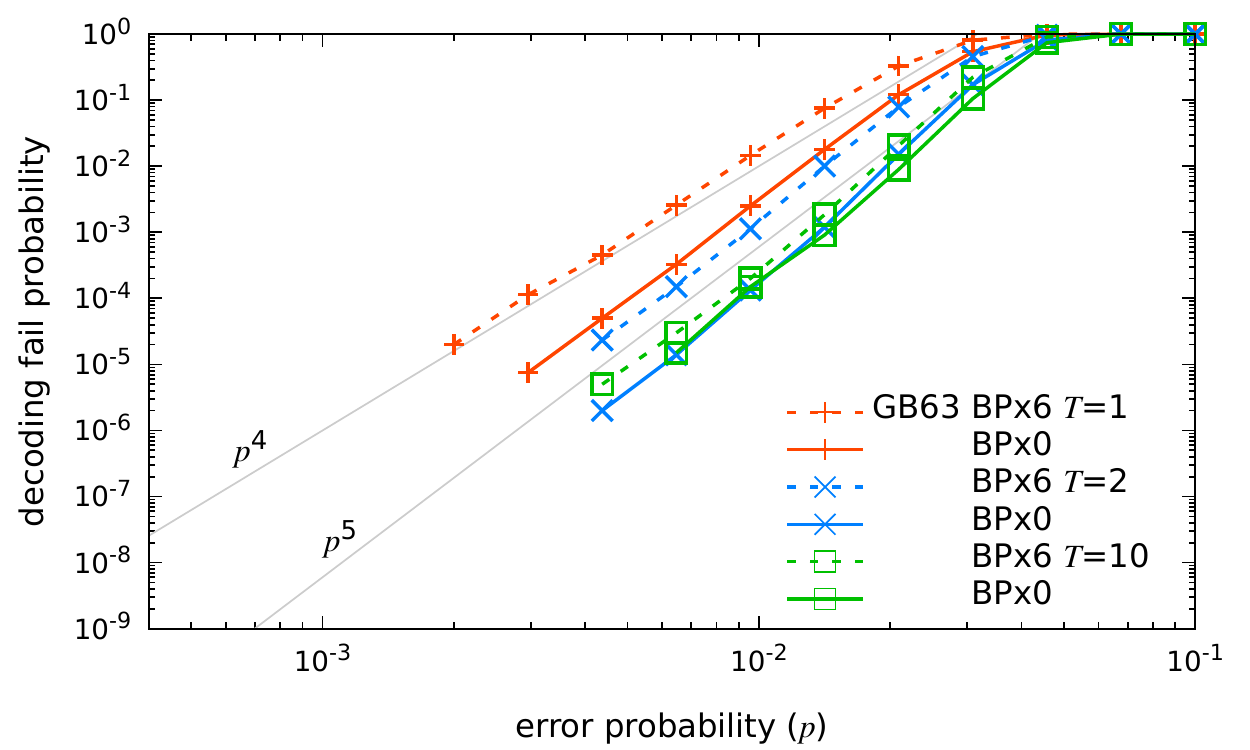}
\end{minipage}
\caption{(Color online) Logical error probability as a function of
  phenomenological model error parameter $p$.  Generalized bicycle
  codes {\tt GB15} $[[30,8,4]]$, {\tt GB31} $[[62,10,6]]$ (plot a),
  and {\tt GB63} $[[126,12,10]]$ (plot b) are labeled by the circulant
  size $\ell$ (see Tab.~\ref{tab:codes}); for a code with distance $d$
  measurements are repeated $N=d$ times.  Parameter $T$ is the
  decoding window size; labels {\tt BPx}$m$ indicate single-step SW
  decoder: $m=0$ with full matrices with all $\ell$ rows, and $m>0$
  with $m$ rows removed (all redundant rows).  Thin black lines
  labeled with powers of $p$ are meant to guide the eye.  See text for
  interpretation of the results.}
\label{fig:Pheno2}
\end{figure}

In Fig.~\ref{fig:Pheno2}(a), logical error probability for sliding
window decoding is plotted as a function of $p$, for codes with
$\ell=15$ ({\tt GB15} $[[30,8,4]]$) and $\ell=31$ ({\tt GB31}
$[[62,10,6]]$), with the size of sliding windows $T$ indicated in the
caption.  Additionally, {\tt BPx}$m$ indicates that $m$ redundant rows
have been removed ($m=0$ corresponds to full matrices with $\ell$
rows, while $m=4$, $5$ or $6$ for codes with $k=8$, $10$, or $12$,
respectively, leaves only linearly independent rows in $H_X$.)

\begin{figure}[htbp]
  \centering
  \raisebox{1.4em}{(a)}\;%
  \includegraphics[width=0.7\linewidth]{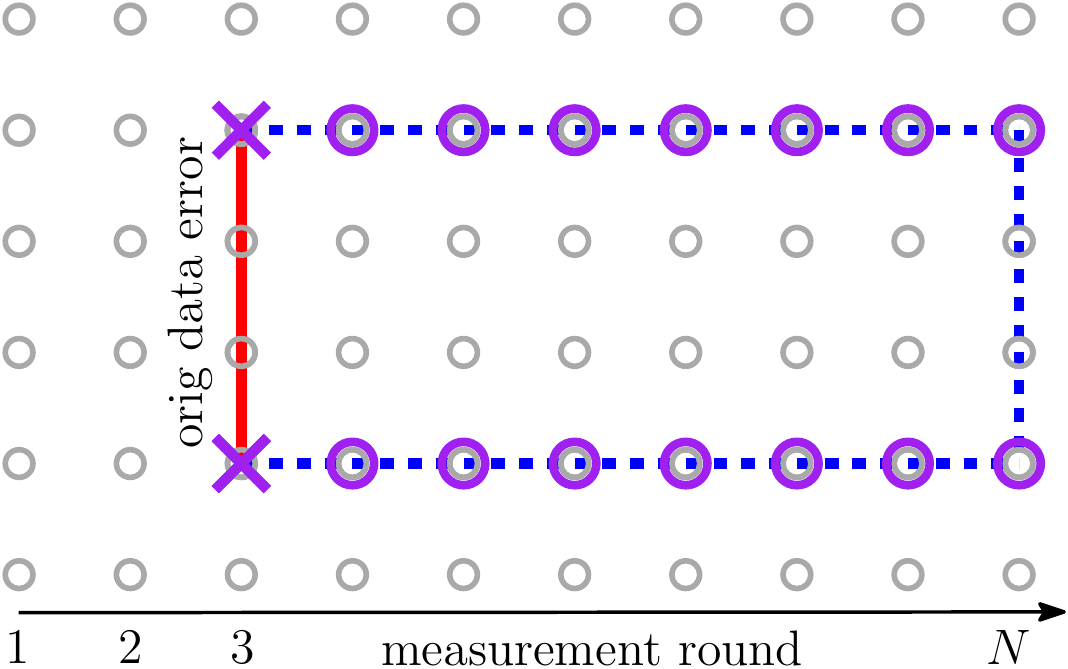}\\[2em]
  \raisebox{0.2em}{(b)}\;%
  \includegraphics[width=0.27\linewidth]{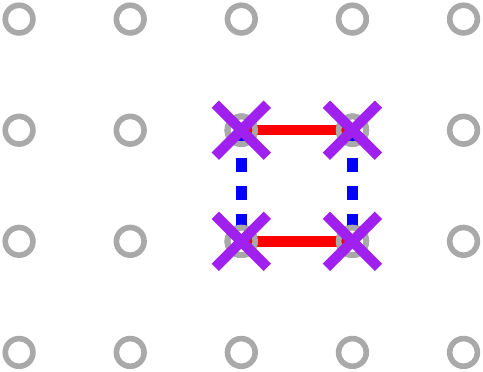}\label{fig:illustrateC}
  \qquad
  \raisebox{0.2em}{(c)}\;%
  \includegraphics[width=0.27\linewidth]{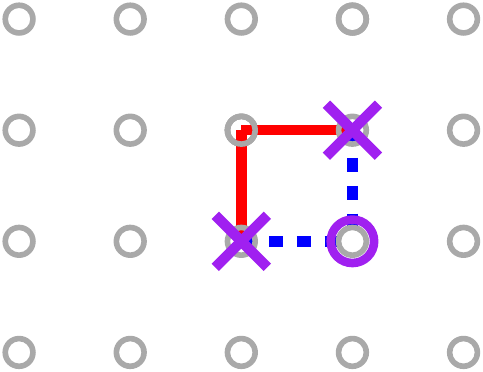}\label{fig:illustrateD}
  \caption{(Color online) Illustration of one-step single-shot
    decoding with phenomenological error model based on a repetition
    code.  Vertical and horizontal edges correspond to possible
    locations for data and measurement errors, respectively.  (a) Red
    solid line in the $3^{\rm rd}$ measurement round shows a weight-3
    data error with weight-2 syndrome (purple $\times$ symbols). The
    $T=1$ SW decoder repeatedly decodes this as a weight-2 measurement
    error (blue dashed lines), which triggers additional detector
    events (purple circles) until the very last round.  The decoded
    error is equivalent to the original one, but long chains violate
    the LDPC property of the code.  In comparison, $T=2$ SW decoder
    recognizes this error correctly as a data error; only a weight-5 data
    error may trigger long error chains running to the last
    measurement round, $t=N$.  (b) A small error bubble is generated
    by a weight-2 measurement error in two neighboring syndrome bits.
    The $T=1$ SW decoder interprets these as weight-1 data errors in
    two consecutive layers.  (c) An error bubble from a weight-1 data
    error and an adjacent measurement error.  The $T=1$ SW decoder
    predicts a measurement error and a data error equivalent to the
    original error.}
\label{fig:1steptoymodel}
\end{figure}

The plots for each code run largely parallel to each other, with
logarithmic slopes consistent with code distances, except for the line
{\tt GB31} {\tt BPx5} $T=1$, that is, single-shot decoding for the
code $[[62,10,6]]$ with all redundant rows removed.  Here
substantially bigger logical error rates are seen, and the slope
becomes quadratic at $p\lesssim 10^{-3}$.  We verified that in this
case some weight-two data errors may be decoded as a weight-one
data and a weight-one syndrome error; the syndrome error propagates to
the last decoding round where it may trigger a different weight-three
data error, resulting in an overall data error of weight six
corresponding to a non-trivial codeword.  See
Fig.~\ref{fig:1steptoymodel} for an illustration of syndrome error
propagation with this decoder.

Fig.~\ref{fig:Pheno2}(b) shows similar data for the code {\tt GB63}
with parameters $[[126,12,10]]$.  These data show much larger
variation, although the logarithmic slope is, again, consistent with
the code distance, except for the single-shot decoding with all
redundant rows removed.  Here, some weight-four errors cannot be
corrected.

\begin{figure}[htbp]
  \centering
  \includegraphics[width=.5\textwidth]{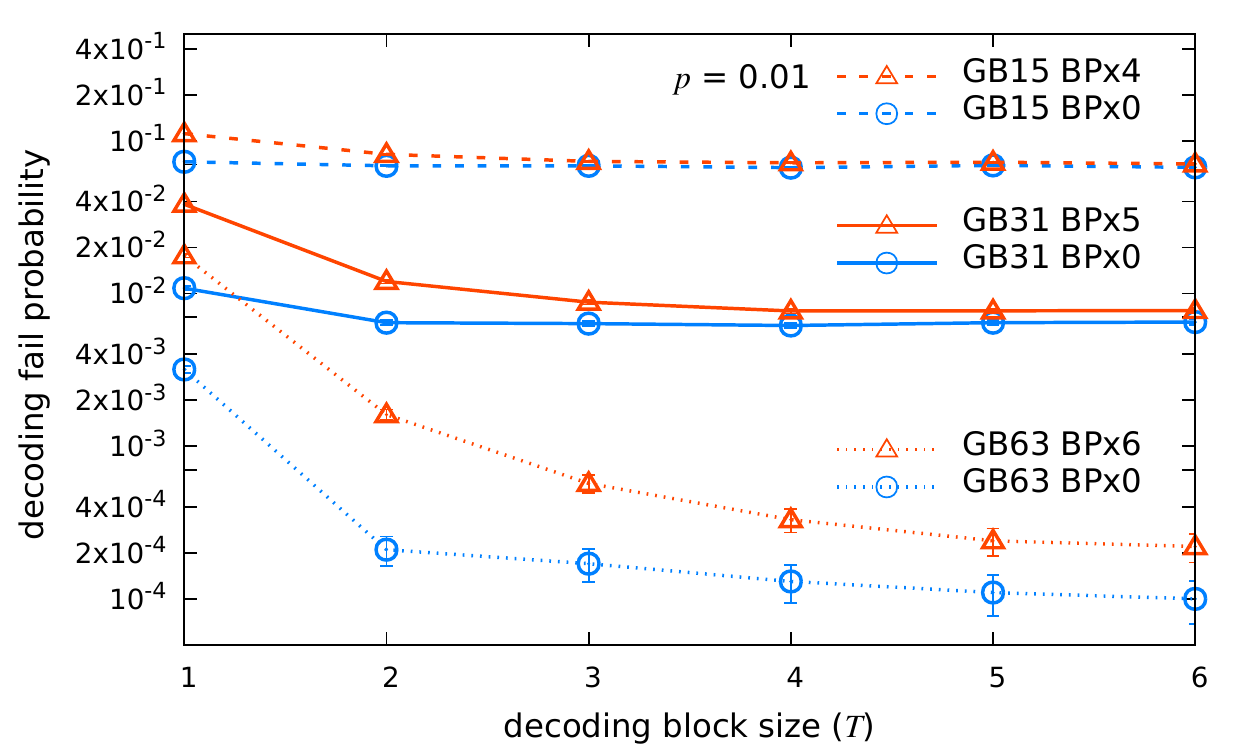}
  \caption{(Color online) Same as Fig.~\ref{fig:Pheno2} as a function
    of SW size $T$, at phenomenological model error probability
    $p=0.01$.  Only results for single-step decoding are shown.  With
    all redundant rows present (blue circles), already 2-shot decoding
    ($T=2$) nearly saturates the logical error probability, except for
    the biggest code where some downward slope is also seen for
    $T>2$.}
\label{fig:Pheno1}
\end{figure}

To visualize the differences between different sequential decoding
protocols, in Fig.~\ref{fig:Pheno1} we show logical error rates at
fixed $p=0.01$ as a function of the sliding window size $T$.  Results
for all three codes are shown, with the number of repeated
measurements equal to the code distance, $N=d$, except for the code
{\tt GB15} where we used $N=6$.  Data points shown in
Fig.~\ref{fig:Pheno1} with blue circles correspond to full matrices,
while red triangles show results for matrices will all redundant rows
removed.  To reduce visual clutter, we do not show simulation results
for matrices with fewer redundant rows removed; generally, fail
probability increases monotonously with the number of removed rows.

\begin{figure}[htbp]
  \raisebox{1.4em}{(a)}\;%
  \includegraphics[width=0.9\linewidth]{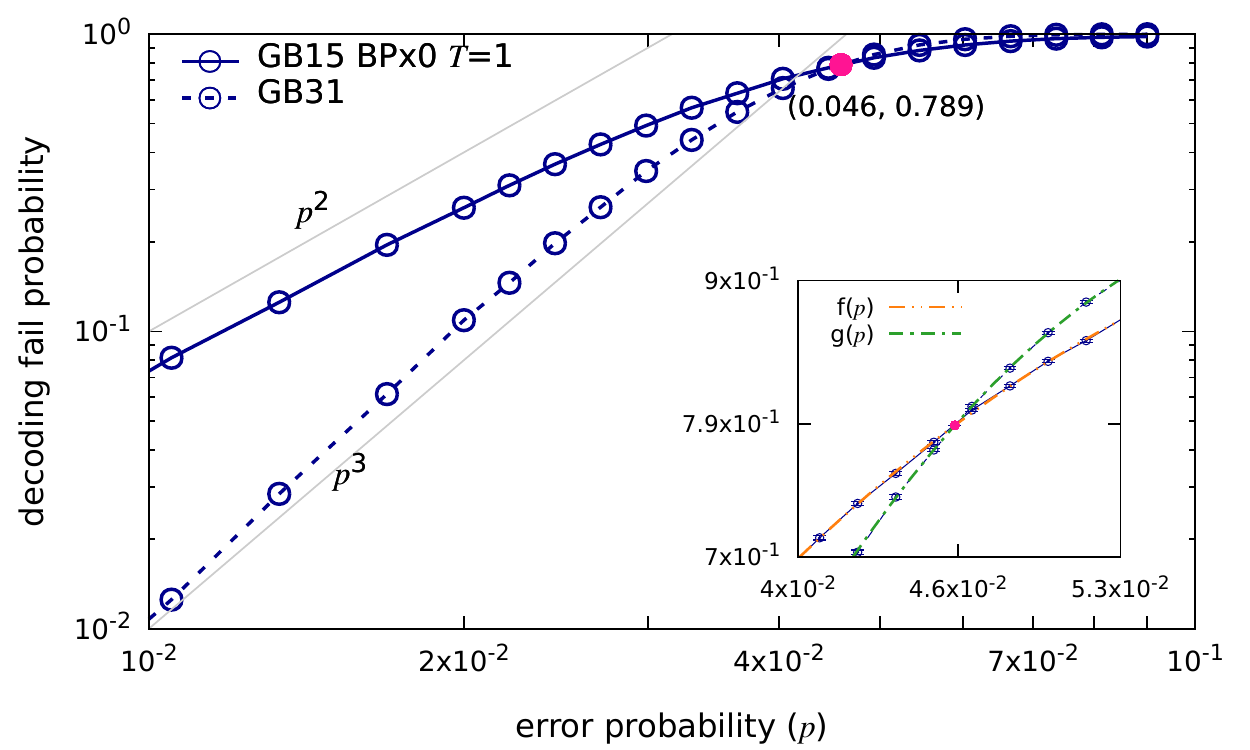}\\
  \raisebox{1.4em}{(b)}\;%
  \includegraphics[width=0.9\linewidth]{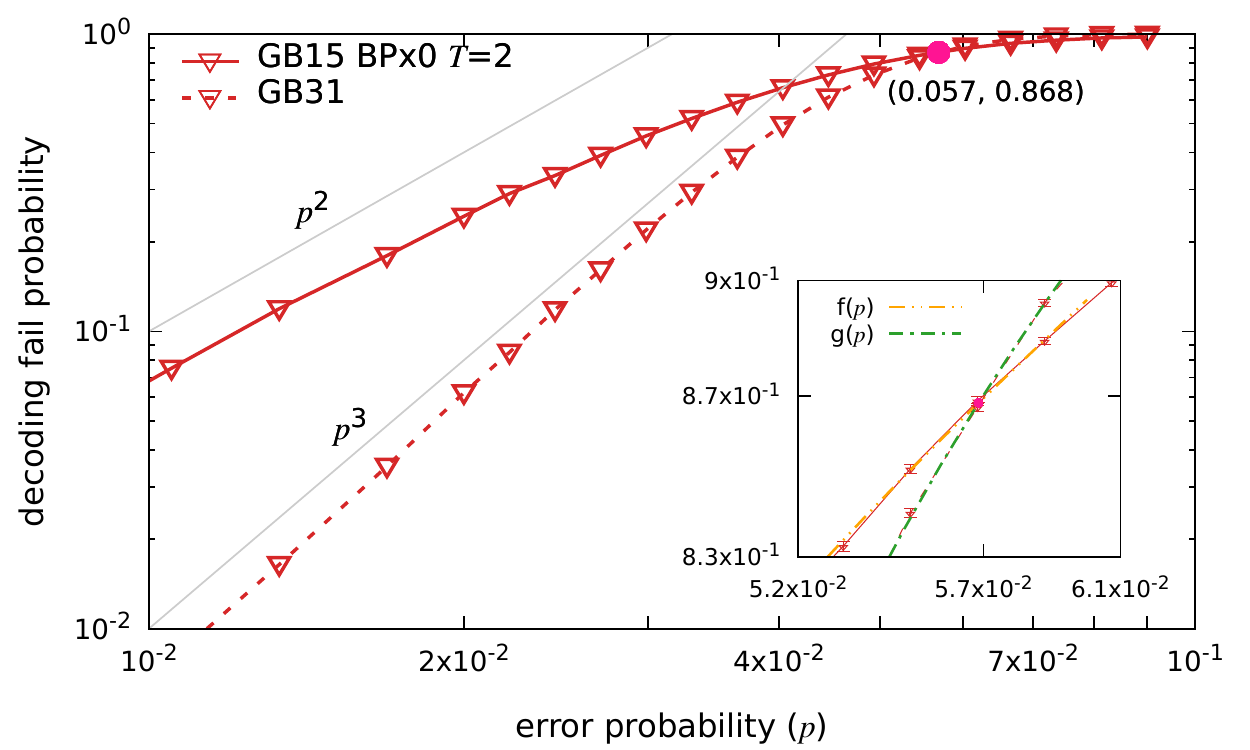}
\label{fig:pstarT2}
\caption{Logical error probability as a function of phenomenological
  error probability $p$ for {\tt GB15} $[[30,8,4]]$ and {\tt GB31}
  $[[62,10,6]]$ codes, with all redundant rows preserved.  Single step
  SW decoder, with (a) $T=1$, (b) $T=2$ window size.  Insets show the
  intersection region in more detail.  The coordinates of intersection
  points give pseudothresholds and the corresponding logical fail
  probabilities.}
\label{fig:threshold-x0}
\end{figure}

\begin{figure}[htbp]
  \includegraphics[width=\linewidth]{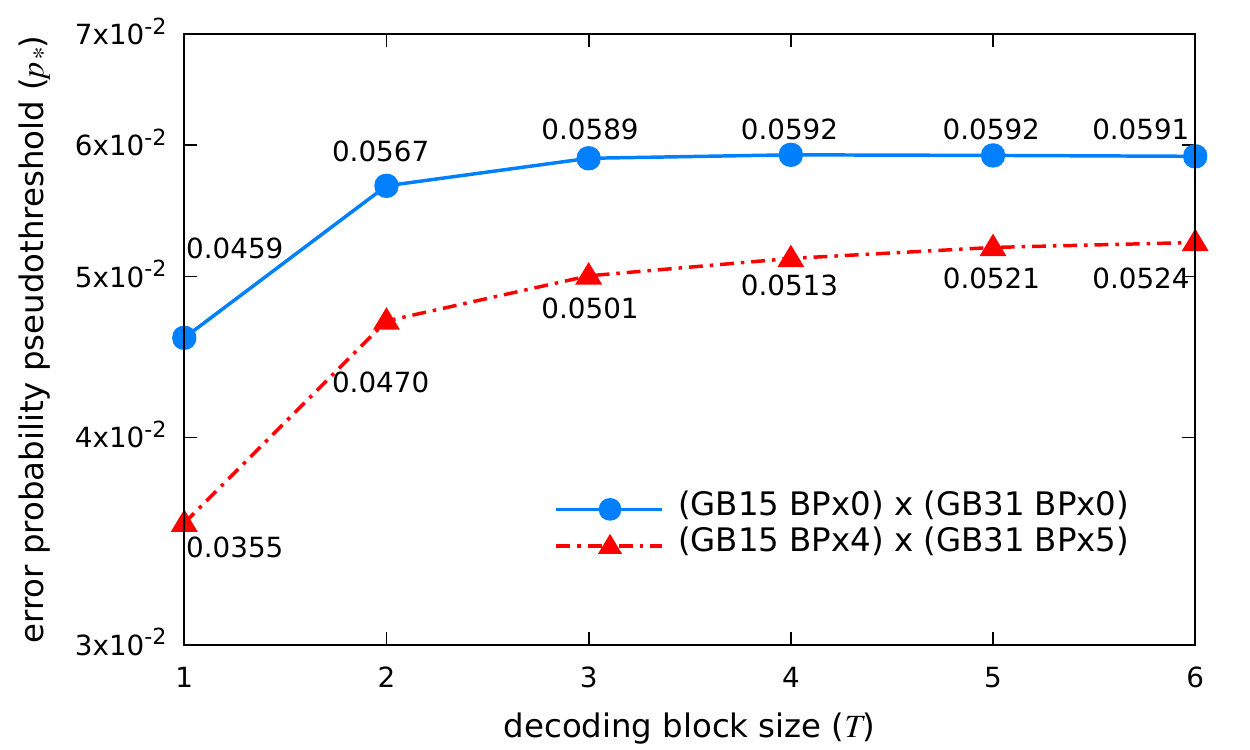}
  \caption{(Color online) Phenomenological error model error
    probability pseudothresholds $p_*$ from the intersection of
    logical error probability curves for {\tt GB15} $[[30,8,4]]$ and
    {\tt GB31} $[[62,10,6]]$ codes, plotted as a function of decoding
    window size $T$.  Blue circles and red triangles, respectively,
    give the results for full-size matrices ({\tt BPx0}) and those
    with redundant rows removed.  Better decoding accuracy gives
    better error tolerance and higher thresholds,
    cf.~Fig.~\ref{fig:Pheno1}.}
\label{fig:pssttarAll}
\end{figure}

As would be expected, tendency for better decoding (and thus better
error tolerance) with increased window size and increased row redundancy
is accompanied by higher thresholds, see Fig.~\ref{fig:threshold-x0}
for a sample of high-resolution threshold-style plots and
Fig.~\ref{fig:pssttarAll} for the plots of (pseudo)threshold
location as a function of decoding window size, separately for the
full-size matrices and those with redundant rows removed.

Our first observation is that, as expected, redundant stabilizer
generators improve decoding accuracy for all decoding window sizes.
Somewhat surprisingly, there is no variation of the exponent (the
slope on log-log plots), except for $T=1$.  Thus, for these codes, in
the case of phenomenological error model, sliding window decoding protocols
with $T\ge 2$ correct the same number of errors as the full-block
decoding.

To illustrate the operation of a single-shot one-step decoder, in
Fig.~\ref{fig:1steptoymodel} we give several examples for a
phenomenological error model based on a large-distance repetition
code.  In Fig.~\ref{fig:1steptoymodel}(a), the original weight-3 error
(vertical red solid line with purple $\bs\times$ symbols at the ends
to indicate the triggered detectors) is repeatedly decoded as a
weight-2 measurement error, which triggers additional detector events
as indicated by purple circles.  This happens until the very last
decoding round, $t=N$, where a data error is the only option.  While
the error is decoded correctly (the long chain is in the same
equivalence class as the original data error), the $T=1$ decoder
clearly has a propensity for generating large-weight error chains,
thus violating the LDPC property of the code.  Unlike for small
bubbles in Fig.~\ref{fig:1steptoymodel}(b) and (c), there is more
likelihood that two or more such long chains together would cause a
logical failure.

In the absence of redundant checks, already a single-bit data error
near a boundary has a 50\% chance to be decoded as a measurement
error, while a $w=2$ data chain there triggers a long chain.  The
likelihood of such long chains is reduced with increased row
redundancy (or syndrome distance) and increased decoding window size
$T$.  Indeed, here bigger-weight data errors are required for a
triggered detector to be incorrectly decoded as a measurement error.

\begin{figure}[htbp]
  \centering
  \includegraphics[width=1.\columnwidth]{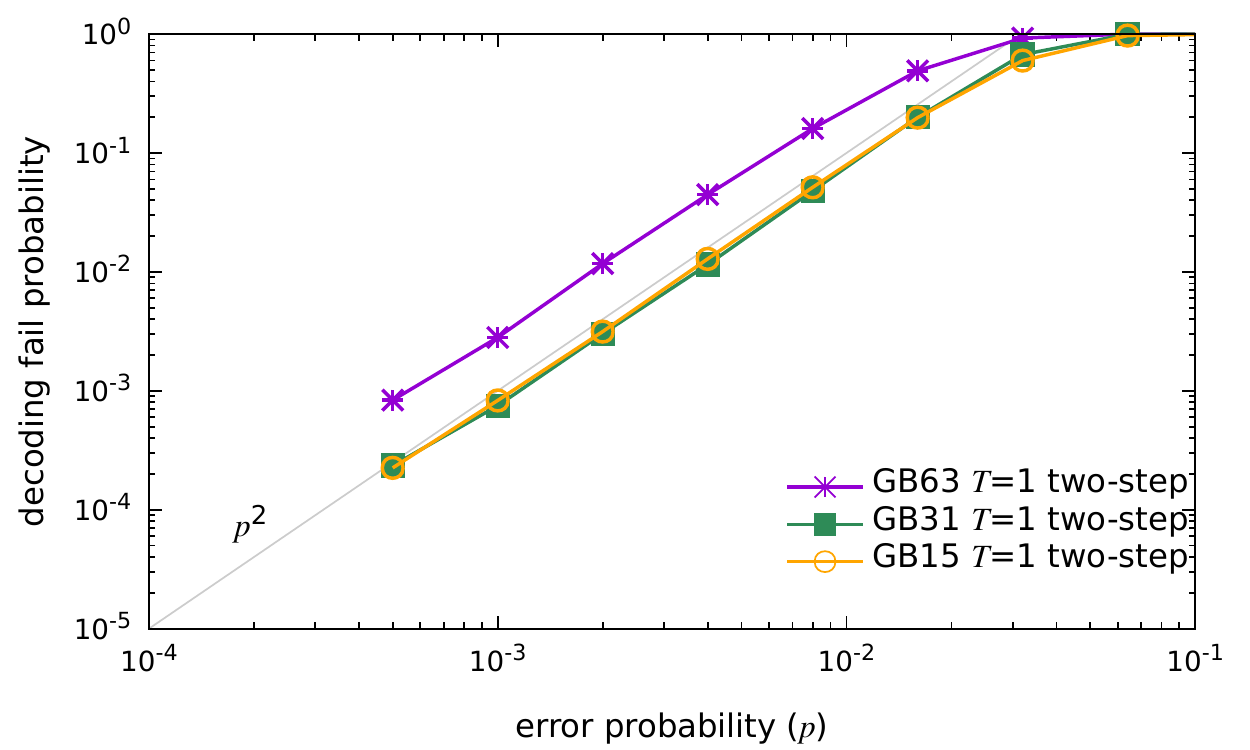}
  \caption{(Color online) As in Fig.~\ref{fig:Pheno2}, but for the
    single-shot two-step decoder, where syndrome and data errors are
    corrected separately.  Only the results for full $\ell$-row
    matrices are give.  For all three codes, {\tt GB15} $[[30,8,4]]$,
    {\tt GB31} $[[62,10,6]]$, and {\tt GB63} $[[126,12,10]]$, some
    weight-two errors cause decoding failures.  This is consistent
    with syndrome distance $d_{\rm S}=3$, since weight-2 syndrome
    errors may result in failed decoding.}
\label{fig:Pheno3}
\end{figure}

For comparison, we have also implemented a single-shot two-step
decoder\cite{Campbell-2018}, where in the first step, the redundancy
between measured syndrome bits is used to correct measurement errors,
followed by a data-decoding step with the corrected syndrome.  The
corresponding data are shown in Fig.~\ref{fig:Pheno3}.  Evidently,
this decoder does not work well here, as the logical error rates are
highest for the code with the largest distance.  Regardless of the
code distance, the plots scale quadratically with $p$, indicating that
weight-2 errors cause decoding failures.

While for $\ell=15$ distance-3 code this is expected, the situation is
different for the longer codes with higher distances.  Both codes have
even distances; there are some weight-2 measurement errors, say, in
round $t=1$, decoded to a weight-3 syndrome corresponding to two
equal-weight errors which add up to a non-trivial codeword.  The same
(corrected) syndrome is seen twice by the data-qubit decoder; in half
of the cases two different errors $\bs e_1\neq\bs e_2$ are produced
resulting in a logical error.  Evidently, the reason for decoder
failure is that the syndrome code does not have good
confinement\cite{Quintavalle-Vasmer-Roffe-Campbell-2021}: a
hight-weight error can give a small-weight syndrome.

Interestingly, the $T=1$ two-step decoder works correctly for the
error in Fig.~\ref{fig:1steptoymodel}(a).  Namely, the data error
would be interpreted as such and corrected immediately, without
generating long chains.  Combined short data/syndrome chains in
Fig.~\ref{fig:1steptoymodel}(b) and (c) are also decoded correctly.

\subsection{Circuit error model}

Given the original code with $\ell\times 2\ell$ CSS matrices and the
desired measurement order for $X$ and $Z$ generators, we used a script
to construct the full-code {\tt Stim} measurement circuit (denoted
{\tt x0}), and also circuits {\tt x1}, {\tt x2}, etc., respectively
with 1, 2, etc.\ redundant rows removed, where gates for the missing
ancillary qubits would be just skipped.  Two variants of each circuit
were generated: one with each data qubit initialized in the $\ket0$
state and measured in the $Z$ basis at the end, and another
initialized in the $\ket+$ state and measured in the $X$ basis.  Each
circuit included data qubit initialization, initial measurement cycle
discarding random measurements ($X$- or $Z$-generators, depending on
the variant), $N-2$ regular measurement cycles, and a final round with
every data qubit measured.  Circuits also included one- and two-qubit
Pauli errors according with the error model, and annotations needed to
convert measurement results to detector events and observables.  {\tt
  Stim} was used both to construct the corresponding DEM files (which
were used to calculate the circuit distances and produce the matrices
associated with the code) and to simulate the circuits.

The sampled data was either sent to {\tt vecdec} for decoding directly
(decoder window size $T=N$), or split at appropriate columns and sent
for decoding sequentially as described in Sec.~\ref{sec:decoding}.
Matrices and error probabilities needed for decoding were extracted
from the DEM file and split into submatrices as needed with the help
of utility programs included with {\tt vecdec}
distribution\cite{Pryadko-2025-vecdec}.

We should emphasize that even relatively short codes constructed here
resulted in rather large DEM matrices.  Specifically, for the largest
code with $n=126$ qubits, with $N=6$ cycles used for circuit
simulations, the circuit error model is equivalent to a binary CSS
code with the check matrix ${\sf H}_X$ of dimensions
$630 \times 23121$.  This makes it difficult both to verify the
circuit distances, and to run BP+OSD decoder, especially given the
fact that OSD is triggered often due to high degeneracy of the circuit
code.  The actual calculations were done on a cluster at the UCR
High-Performance Computing Center.

\begin{figure}[htbp]
  \centering
  \raisebox{1.4em}{(a)}\;%
  \includegraphics[width=0.97\linewidth]{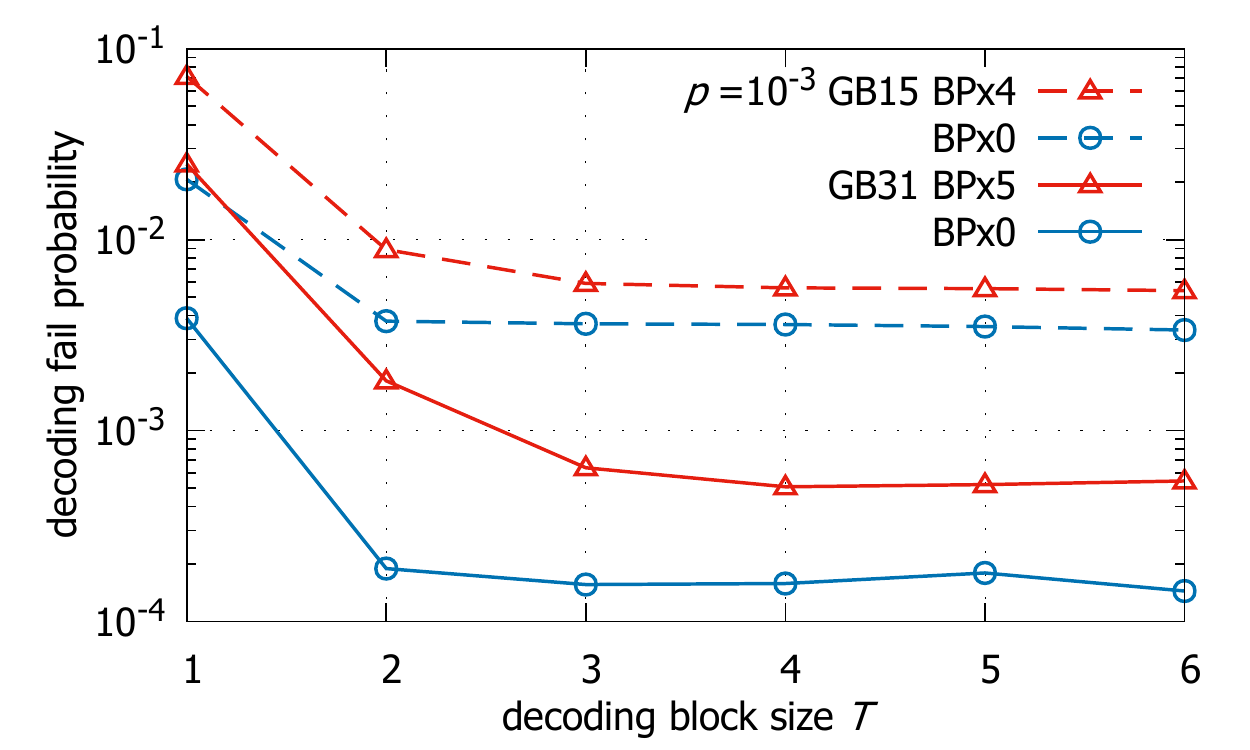}\\
  \raisebox{1.4em}{(b)}\;%
  \includegraphics[width=0.97\linewidth]{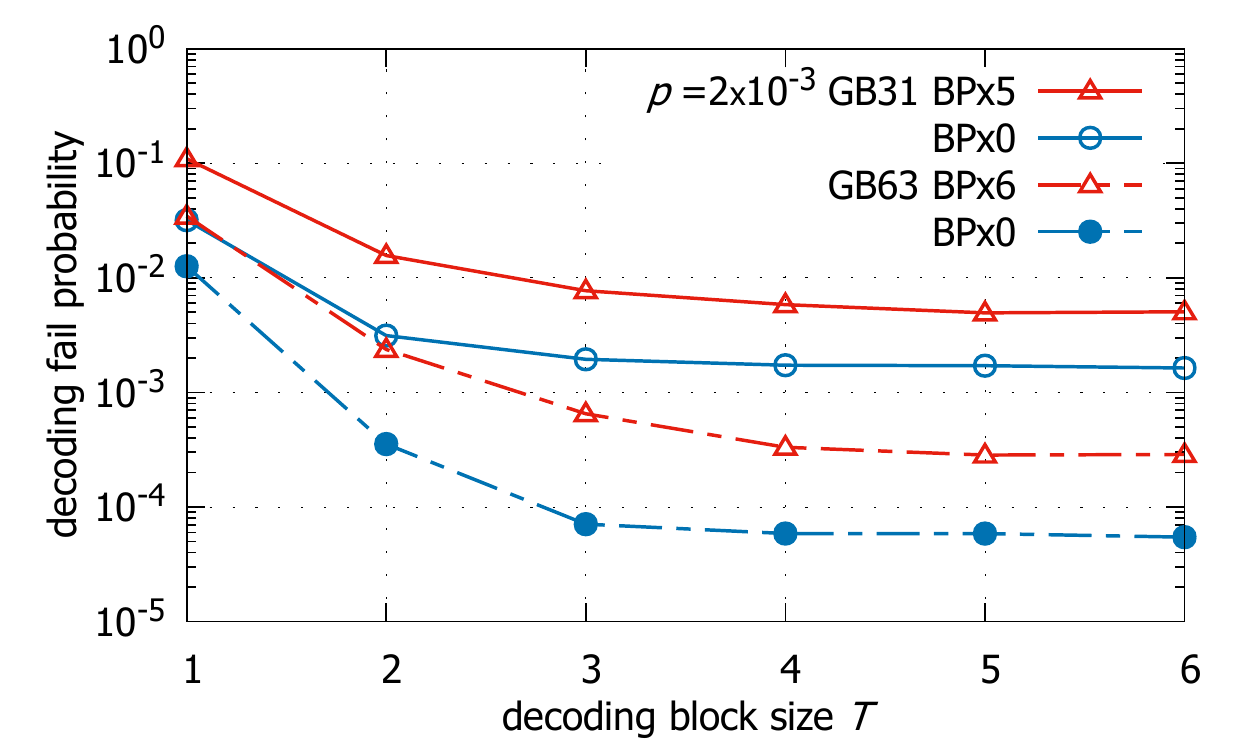}
  \caption{(Color online) Same as Fig.~\ref{fig:Pheno1} but for the
    circuit error model simulations.  Results for codes {\tt GB15}
    $[[30,8,4]]$ and {\tt GB31} $[[62,10,6]]$ at $p=10^{-3}$ are shown
    in (a), and {\tt GB31} $[[62,10,6]]$ and {\tt GB63}
    $[[126,12,10]]$ at $p=2\times 10^{-3}$ in (b).  Here each circuit
    included a fixed number $N=6$ of measurement rounds.}
  \label{fig:circ-new}
\end{figure}

The results for the circuit error model are shown in
Fig.~\ref{fig:circ-new}.  Here logical error rate is shown as a
function of the decoding window size $T$, for codes {\tt GB15} and
{\tt GB31} at $p=10^{-3}$ in Fig.~\ref{fig:circ-new}(a) and {\tt GB31}
and {\tt GB63} at $p=2\times 10^{-3}$ in Fig.~\ref{fig:circ-new}(b).
Blue circles show the results with full-size matrices used in the
circuits, and red triangles with redundant rows dropped.

For the parameters used, redundant syndrome bits give over an order of
magnitude reduction of logical error rates, even at larger window
sizes $T$.  In addition, data with full-size matrices show better
convergence at small $T$: two-shot decoding ($T=2$) is sufficient for
the two shorter codes and the error rate used in
Fig.~\ref{fig:circ-new}(a), while $T=3$ is required for optimal
accuracy for the codes and error rate in Fig.~\ref{fig:circ-new}(b).

In comparison, same codes without redundant generators require larger
decoding window sizes and show saturation at much higher logical error
rates.  In particular, some weight-one errors occasionally are not
decoded correctly with a $T=1$ SW decoder.  Namely, some single-bit
data errors may be decoded as a single-bit measurement error and a
single-bit data error, with the syndrome-bit error eventually inducing
a high-weight data error, which results in a logical error.  This
happens because BP decoder is not a true minimum-weight decoder; the
convergence is especially bad in the case of circuit error models due
to the abundance of short cycles.

\section{Conclusions.}

Two-block codes have natural redundancy of smallest-weight stabilizer
generators.  We explored how this redundancy can be used to improve
the decoding in a fault-tolerant regime, i.e., in the presence of
measurement errors, focusing on a sequential ``few-shot'' decoding.
To this end, we constructed a large number of planar GB codes similar
to BB codes designed by Bravyi et al.\cite{Bravyi-etal-Yoder-2023},
but with the maximum syndrome distance, $d_{\rm S}=3$.  While code
parameters are generally different from those constructed earlier
(without regard to syndrome distance), they are similar.  Larger
syndrome distance guarantees optimal decoder accuracy for this class
of codes, especially at smaller window sizes, and increases the
threshold for data errors being incorrectly decoded as measurement
errors.  While the reduction of the decoding fail rates due to
additional redundant rows is just a constant factor under an order of
magnitude, it comes essentially for free.  Use of such codes can
reduce both the decoding complexity and error correction latency due
to faster convergence with the increased window sizes.

Of course, there are other code families with higher thresholds, e.g.,
recently discovered designed for stability to measurement errors
rotated 4-dimensional toric
codes\cite{Aasen-Haah-Hastings-Wang-2025,Aasen-etal-Svore-2025} and
more general abelian multi-cycle
codes\cite{Lin-Lim-Kovalev-Pryadko-2025}, or the finite-rate
concatenated
codes\cite{Yamasaki-Koashi-2024,Yoshida-Tamiya-Yamasaki-2024}.
However, such codes do not have the advantage of a natural planar
layout, and in general require much more connectivity between the
qubits.

\begin{acknowledgments} This work was supported in part by the APS
  M. Hildred Blewett Fellowship (HKL) and the NSF Division of Physics
  via the grant 2112848 (LPP).
\end{acknowledgments}

\appendix
\section{All constructed codes}
\begin{longtable}[c]{c|c|c|c|c|c|c|c}
  $n$&$k$&$d$&$d_{\rm S}$&$n_X$&$n_Y$&$a$&$b$\\ \hline
30 & 8 & 4 & 3 & 15 & 1 & $1+x+x^{4}$ & $1+x^{2}+x^{8}$ \\ 
42 & 10 & 4 & 3 & 21 & 1 & $1+x+x^{5}$ & $1+x^{2}+x^{10}$ \\ 
56 & 12 & 4 & 3 & 14 & 2 & $1+x+x^{3}y$ & $1+x^{2}+x^{6}$ \\ 
60 & 16 & 4 & 3 & 30 & 1 & $1+x^{2}+x^{8}$ & $1+x^{4}+x^{16}$ \\ 
62 & 10 & 4 & 3 & 31 & 1 & $1+x+x^{12}$ & $1+x^{2}+x^{24}$ \\ 
62 & 10 & 6 & 3 & 31 & 1 & $1+x+x^{12}$ & $1+x^{3}+x^{8}$ \\ 
72 & 12 & 6 & 3 & 6 & 6 & $1+y+x^{3}y^{2}$ & $1+xy+x^{5}y^{2}$ \\ 
84 & 20 & 4 & 3 & 42 & 1 & $1+x^{2}+x^{10}$ & $1+x^{4}+x^{20}$ \\ 
90 & 12 & 4 & 3 & 15 & 3 & $1+x+x^{2}y$ & $1+xy^{2}+x^{8}$ \\ 
90 & 12 & 6 & 3 & 15 & 3 & $1+x+x^{2}y$ & $1+x^{4}y+x^{11}$ \\ 
90 & 24 & 4 & 3 & 15 & 3 & $1+x+x^{4}$ & $1+x^{2}+x^{8}$ \\ 
90 & 24 & 4 & 3 & 45 & 1 & $1+x^{3}+x^{12}$ & $1+x^{6}+x^{24}$ \\ 
96 & 12 & 6 & 3 & 12 & 4 & $1+x+x^{2}y$ & $1+xy^{3}+x^{5}$ \\ 
96 & 16 & 4 & 3 & 12 & 4 & $1+x+x^{2}y$ & $1+x^{2}y^{2}+x^{10}$ \\ 
98 & 12 & 4 & 3 & 7 & 7 & $1+y+x$ & $1+y^{2}+x^{2}$ \\ 
98 & 18 & 4 & 3 & 7 & 7 & $1+x+x^{3}$ & $1+y+y^{3}$ \\ 
112 & 24 & 4 & 3 & 14 & 4 & $1+x+x^{3}y^{2}$ & $1+x^{2}+x^{6}$ \\ 
112 & 24 & 4 & 3 & 28 & 2 & $1+x^{2}+x^{6}y$ & $1+x^{4}+x^{12}$ \\ 
120 & 16 & 4 & 3 & 10 & 6 & $1+x+x^{3}y$ & $1+x^{2}+x^{6}y^{2}$ \\ 
120 & 16 & 4 & 3 & 30 & 2 & $1+x+x^{4}y$ & $1+x^{2}+x^{8}$ \\ 
120 & 16 & 6 & 3 & 10 & 6 & $1+y+x^{5}y^{2}$ & $1+x^{2}y^{2}+x^{4}$ \\ 
120 & 16 & 6 & 3 & 30 & 2 & $1+x+x^{4}y$ & $1+x^{4}+x^{16}$ \\ 
120 & 32 & 4 & 3 & 10 & 6 & $1+x^{2}+x^{6}y^{2}$ & $1+x^{2}y^{2}+x^{8}y^{2}$ \\ 
120 & 32 & 4 & 3 & 30 & 2 & $1+x^{2}+x^{8}$ & $1+x^{4}+x^{16}$ \\ 
120 & 32 & 4 & 3 & 60 & 1 & $1+x^{4}+x^{16}$ & $1+x^{8}+x^{32}$ \\ 
124 & 20 & 4 & 3 & 62 & 1 & $1+x^{2}+x^{24}$ & $1+x^{4}+x^{48}$ \\ 
124 & 20 & 6 & 3 & 62 & 1 & $1+x^{2}+x^{24}$ & $1+x^{6}+x^{16}$ \\ 
126 & 12 & 10 & 3 & 63 & 1 & $1+x+x^{6}$ & $1+x^{11}+x^{25}$ \\ 
126 & 12 & 4 & 3 & 63 & 1 & $1+x+x^{6}$ & $1+x^{2}+x^{12}$ \\ 
126 & 12 & 6 & 3 & 63 & 1 & $1+x+x^{6}$ & $1+x^{7}+x^{26}$ \\ 
126 & 12 & 8 & 3 & 63 & 1 & $1+x+x^{6}$ & $1+x^{4}+x^{24}$ \\ 
126 & 14 & 6 & 3 & 21 & 3 & $1+xy+x^{5}y^{2}$ & $1+x+x^{5}$ \\ 
126 & 16 & 4 & 3 & 63 & 1 & $1+x+x^{8}$ & $1+x^{2}+x^{16}$ \\ 
146 & 18 & 4 & 3 & 73 & 1 & $1+x+x^{9}$ & $1+x^{2}+x^{18}$ \\ 
150 & 16 & 8 & 3 & 5 & 15 & $1+x+x^{3}y^{5}$ & $1+y+y^{4}$ \\ 
170 & 16 & 10 & 3 & 85 & 1 & $1+x+x^{16}$ & $1+x^{4}+x^{64}$ \\ 
170 & 16 & 4 & 3 & 85 & 1 & $1+x+x^{16}$ & $1+x^{2}+x^{32}$ \\ 
186 & 14 & 10 & 3 & 93 & 1 & $1+x+x^{14}$ & $1+x^{4}+x^{56}$ \\ 
186 & 14 & 4 & 3 & 93 & 1 & $1+x+x^{14}$ & $1+x^{2}+x^{28}$ \\ 
186 & 14 & 6 & 3 & 93 & 1 & $1+x+x^{14}$ & $1+x^{17}+x^{49}$ \\ 
210 & 14 & 10 & 3 & 105 & 1 & $1+x+x^{12}$ & $1+x^{3}+x^{79}$ \\ 
210 & 14 & 12 & 3 & 105 & 1 & $1+x+x^{12}$ & $1+x^{16}+x^{87}$ \\ 
210 & 14 & 4 & 3 & 105 & 1 & $1+x+x^{12}$ & $1+x^{2}+x^{24}$ \\ 
210 & 14 & 8 & 3 & 105 & 1 & $1+x+x^{12}$ & $1+x^{8}+x^{96}$ \\ 
288 & 16 & 12 & 3 & 12 & 12 & $1+x+x^{2}y^{3}$ & $1+y+x^{3}y^{8}$ \\ 
288 & 20 & 6 & 3 & 12 & 12 & $1+x+x^{5}y^{3}$ & $1+y+x^{3}y^{5}$ \\ 
294 & 18 & 10 & 3 & 7 & 21 & $1+x+x^{3}y^{3}$ & $1+y+y^{17}$ \\ 
294 & 30 & 4 & 3 & 7 & 21 & $1+x+x^{3}$ & $1+y+y^{5}$ \\ 
392 & 18 & 12 & 3 & 14 & 14 & $1+x+x^{2}y^{5}$ & $1+y+x^{7}y^{5}$ \\ 
392 & 24 & 12 & 3 & 14 & 14 & $1+x+x^{3}y^{7}$ & $1+y+x^{7}y^{5}$ \\ 
450 & 16 & 16 & 3 & 15 & 15 & $1+x+x^{2}y^{2}$ & $1+y+x^{5}y^{4}$ \\ 
450 & 16 & 18 & 3 & 15 & 15 & $1+x+x^{3}y$ & $1+y+x^{8}y^{11}$ \\ 
450 & 16 & 20 & 3 & 15 & 15 & $1+x+x^{2}y^{4}$ & $1+y+x^{5}y^{9}$ \\ 
450 & 16 & 8 & 3 & 15 & 15 & $1+x+x^{2}y$ & $1+y+y^{4}$ \\ 
450 & 20 & 10 & 3 & 15 & 15 & $1+x+x^{3}y^{5}$ & $1+y+x^{10}y^{4}$ \\ 
450 & 20 & 14 & 3 & 15 & 15 & $1+x+x^{3}y^{5}$ & $1+y+x^{5}y^{12}$ \\ 
450 & 20 & 16 & 3 & 15 & 15 & $1+x+x^{4}y$ & $1+y+xy^{10}$ \\ 
450 & 24 & 8 & 3 & 15 & 15 & $1+x+x^{3}y^{5}$ & $1+y+y^{4}$ \\ 
450 & 28 & 10 & 3 & 15 & 15 & $1+x+y^{4}$ & $1+y+x^{4}$ \\ 
450 & 28 & 4 & 3 & 15 & 15 & $1+x+y^{2}$ & $1+y+x^{8}$ \\ 
450 & 32 & 8 & 3 & 15 & 15 & $1+x+x^{4}$ & $1+y+y^{4}$ \\ 
882 & 20 & 10 & 3 & 21 & 21 & $1+x+x^{2}y^{3}$ & $1+y+y^{5}$ \\ 
882 & 20 & 12 & 3 & 21 & 21 & $1+x+x^{8}y^{3}$ & $1+y+x^{12}y^{8}$ \\ 
882 & 20 & 6 & 3 & 21 & 21 & $1+x+x^{8}y^{6}$ & $1+y+x^{6}y^{8}$ \\ 
882 & 26 & 10 & 3 & 21 & 21 & $1+x+x^{5}$ & $1+y+x^{9}y^{5}$ \\ 
882 & 50 & 10 & 3 & 21 & 21 & $1+x+x^{5}$ & $1+y+y^{5}$ \\ 
1152 & 24 & 18 & 3 & 24 & 24 & $1+x+x^{5}y^{3}$ & $1+y+x^{15}y^{5}$ \\ 
1152 & 36 & 6 & 3 & 24 & 24 & $1+x+x^{8}y^{9}$ & $1+y+x^{9}y^{8}$ \\ 
\caption{$^{\strut}$Parameters of non-trivial 2BGA codes with distances
    $d>2$ and syndrome distances $d_{\rm S}=3$ over one- and
    two-generator abelian groups, constructed
    from polynomials of weights $w_a=w_b=3$.  Parameters $n_X$ and
    $n_Y$ are the orders of generators $x$ and $y$ in the group
    presentation $\langle x,y| x^{N_X}=y^{N_Y}=xyx^{-1}y^{-1}=1\rangle$;
    $n_Y=1$ gives a cyclic group $C_{n_X}$.
    All such codes with distinct parameters and group orders $\ell=n_Xn_Y\le
    75$ are listed, and some of the codes for larger groups.  
  }
  \label{tab:big-codes}
\end{longtable}

\bibliography{lpp,qc_all,more_qc,xr,ldpc}

\begin{thebibliography}{67}%
\makeatletter
\providecommand \@ifxundefined [1]{%
 \@ifx{#1\undefined}
}%
\providecommand \@ifnum [1]{%
 \ifnum #1\expandafter \@firstoftwo
 \else \expandafter \@secondoftwo
 \fi
}%
\providecommand \@ifx [1]{%
 \ifx #1\expandafter \@firstoftwo
 \else \expandafter \@secondoftwo
 \fi
}%
\providecommand \natexlab [1]{#1}%
\providecommand \enquote  [1]{``#1''}%
\providecommand \bibnamefont  [1]{#1}%
\providecommand \bibfnamefont [1]{#1}%
\providecommand \citenamefont [1]{#1}%
\providecommand \href@noop [0]{\@secondoftwo}%
\providecommand \href [0]{\begingroup \@sanitize@url \@href}%
\providecommand \@href[1]{\@@startlink{#1}\@@href}%
\providecommand \@@href[1]{\endgroup#1\@@endlink}%
\providecommand \@sanitize@url [0]{\catcode `\\12\catcode `\$12\catcode
  `\&12\catcode `\#12\catcode `\^12\catcode `\_12\catcode `\%12\relax}%
\providecommand \@@startlink[1]{}%
\providecommand \@@endlink[0]{}%
\providecommand \url  [0]{\begingroup\@sanitize@url \@url }%
\providecommand \@url [1]{\endgroup\@href {#1}{\urlprefix }}%
\providecommand \urlprefix  [0]{URL }%
\providecommand \Eprint [0]{\href }%
\providecommand \doibase [0]{https://doi.org/}%
\providecommand \selectlanguage [0]{\@gobble}%
\providecommand \bibinfo  [0]{\@secondoftwo}%
\providecommand \bibfield  [0]{\@secondoftwo}%
\providecommand \translation [1]{[#1]}%
\providecommand \BibitemOpen [0]{}%
\providecommand \bibitemStop [0]{}%
\providecommand \bibitemNoStop [0]{.\EOS\space}%
\providecommand \EOS [0]{\spacefactor3000\relax}%
\providecommand \BibitemShut  [1]{\csname bibitem#1\endcsname}%
\let\auto@bib@innerbib\@empty
\bibitem [{\citenamefont {Dennis}\ \emph {et~al.}(2002)\citenamefont {Dennis},
  \citenamefont {Kitaev}, \citenamefont {Landahl},\ and\ \citenamefont
  {Preskill}}]{Dennis-Kitaev-Landahl-Preskill-2002}%
  \BibitemOpen
  \bibfield  {author} {\bibinfo {author} {\bibfnamefont {E.}~\bibnamefont
  {Dennis}}, \bibinfo {author} {\bibfnamefont {A.}~\bibnamefont {Kitaev}},
  \bibinfo {author} {\bibfnamefont {A.}~\bibnamefont {Landahl}},\ and\ \bibinfo
  {author} {\bibfnamefont {J.}~\bibnamefont {Preskill}},\ }\bibfield  {title}
  {\bibinfo {title} {Topological quantum memory},\ }\href
  {https://doi.org/10.1063/1.1499754} {\bibfield  {journal} {\bibinfo
  {journal} {J. Math. Phys.}\ }\textbf {\bibinfo {volume} {43}},\ \bibinfo
  {pages} {4452} (\bibinfo {year} {2002})}\BibitemShut {NoStop}%
\bibitem [{\citenamefont {Gong}\ \emph {et~al.}(2024)\citenamefont {Gong},
  \citenamefont {Cammerer},\ and\ \citenamefont
  {Renes}}]{Gong-Cammerer-Renes-2024}%
  \BibitemOpen
  \bibfield  {author} {\bibinfo {author} {\bibfnamefont {A.}~\bibnamefont
  {Gong}}, \bibinfo {author} {\bibfnamefont {S.}~\bibnamefont {Cammerer}},\
  and\ \bibinfo {author} {\bibfnamefont {J.~M.}\ \bibnamefont {Renes}},\
  }\bibfield  {title} {\bibinfo {title} {Toward low-latency iterative decoding
  of qldpc codes under circuit-level noise},\ }\Eprint
  {https://arxiv.org/abs/2403.18901} {arXiv:2403.18901 [quant-ph]}  (\bibinfo
  {year} {2024}),\ \bibinfo {note} {unpublished}\BibitemShut {NoStop}%
\bibitem [{\citenamefont {Skoric}\ \emph {et~al.}(2023)\citenamefont {Skoric},
  \citenamefont {Browne}, \citenamefont {Barnes}, \citenamefont {Gillespie},\
  and\ \citenamefont
  {Campbell}}]{Skoric-Browne-Barnes-Gillespie-Campbell-2023}%
  \BibitemOpen
  \bibfield  {author} {\bibinfo {author} {\bibfnamefont {L.}~\bibnamefont
  {Skoric}}, \bibinfo {author} {\bibfnamefont {D.~E.}\ \bibnamefont {Browne}},
  \bibinfo {author} {\bibfnamefont {K.~M.}\ \bibnamefont {Barnes}}, \bibinfo
  {author} {\bibfnamefont {N.~I.}\ \bibnamefont {Gillespie}},\ and\ \bibinfo
  {author} {\bibfnamefont {E.~T.}\ \bibnamefont {Campbell}},\ }\bibfield
  {title} {\bibinfo {title} {Parallel window decoding enables scalable fault
  tolerant quantum computation},\ }\href
  {https://doi.org/10.1038/s41467-023-42482-1} {\bibfield  {journal} {\bibinfo
  {journal} {Nature Communications}\ }\textbf {\bibinfo {volume} {14}},\
  \bibinfo {pages} {7040} (\bibinfo {year} {2023})}\BibitemShut {NoStop}%
\bibitem [{\citenamefont {Bomb\'{\i}n}(2015)}]{Bombin-2015}%
  \BibitemOpen
  \bibfield  {author} {\bibinfo {author} {\bibfnamefont {H.}~\bibnamefont
  {Bomb\'{\i}n}},\ }\bibfield  {title} {\bibinfo {title} {Single-shot
  fault-tolerant quantum error correction},\ }\href
  {https://doi.org/10.1103/PhysRevX.5.031043} {\bibfield  {journal} {\bibinfo
  {journal} {Phys. Rev. X}\ }\textbf {\bibinfo {volume} {5}},\ \bibinfo {pages}
  {031043} (\bibinfo {year} {2015})}\BibitemShut {NoStop}%
\bibitem [{\citenamefont {Brown}\ \emph {et~al.}(2016)\citenamefont {Brown},
  \citenamefont {Nickerson},\ and\ \citenamefont
  {Browne}}]{Brown-Nickerson-Browne-2016}%
  \BibitemOpen
  \bibfield  {author} {\bibinfo {author} {\bibfnamefont {B.~J.}\ \bibnamefont
  {Brown}}, \bibinfo {author} {\bibfnamefont {N.~H.}\ \bibnamefont
  {Nickerson}},\ and\ \bibinfo {author} {\bibfnamefont {D.~E.}\ \bibnamefont
  {Browne}},\ }\bibfield  {title} {\bibinfo {title} {Fault-tolerant error
  correction with the gauge color code},\ }\href
  {https://doi.org/10.1038/ncomms12302} {\bibfield  {journal} {\bibinfo
  {journal} {Nature Communications}\ }\textbf {\bibinfo {volume} {7}},\
  \bibinfo {pages} {12302} (\bibinfo {year} {2016})}\BibitemShut {NoStop}%
\bibitem [{\citenamefont {Campbell}(2019)}]{Campbell-2018}%
  \BibitemOpen
  \bibfield  {author} {\bibinfo {author} {\bibfnamefont {E.~T.}\ \bibnamefont
  {Campbell}},\ }\bibfield  {title} {\bibinfo {title} {A theory of single-shot
  error correction for adversarial noise},\ }\href
  {https://doi.org/10.1088/2058-9565/aafc8f} {\bibfield  {journal} {\bibinfo
  {journal} {Quantum Science and Technology}\ }\textbf {\bibinfo {volume}
  {4}},\ \bibinfo {pages} {025006} (\bibinfo {year} {2019})},\ \Eprint
  {https://arxiv.org/abs/1805.09271} {1805.09271} \BibitemShut {NoStop}%
\bibitem [{\citenamefont {Fujiwara}(2014)}]{Fujiwara-2014}%
  \BibitemOpen
  \bibfield  {author} {\bibinfo {author} {\bibfnamefont {Y.}~\bibnamefont
  {Fujiwara}},\ }\bibfield  {title} {\bibinfo {title} {Ability of stabilizer
  quantum error correction to protect itself from its own imperfection},\
  }\href {https://doi.org/10.1103/PhysRevA.90.062304} {\bibfield  {journal}
  {\bibinfo  {journal} {Phys. Rev. A}\ }\textbf {\bibinfo {volume} {90}},\
  \bibinfo {pages} {062304} (\bibinfo {year} {2014})}\BibitemShut {NoStop}%
\bibitem [{\citenamefont {Ashikhmin}\ \emph {et~al.}(2014)\citenamefont
  {Ashikhmin}, \citenamefont {Lai},\ and\ \citenamefont
  {Brun}}]{Ashikhmin-Lai-Brun-2014}%
  \BibitemOpen
  \bibfield  {author} {\bibinfo {author} {\bibfnamefont {A.}~\bibnamefont
  {Ashikhmin}}, \bibinfo {author} {\bibfnamefont {C.~Y.}\ \bibnamefont {Lai}},\
  and\ \bibinfo {author} {\bibfnamefont {T.~A.}\ \bibnamefont {Brun}},\
  }\bibfield  {title} {\bibinfo {title} {Robust quantum error syndrome
  extraction by classical coding},\ }in\ \href
  {https://doi.org/10.1109/ISIT.2014.6874892} {\emph {\bibinfo {booktitle}
  {2014 {IEEE} International Symposium on Information Theory}}}\ (\bibinfo
  {year} {2014})\ pp.\ \bibinfo {pages} {546--550}\BibitemShut {NoStop}%
\bibitem [{\citenamefont {Ashikhmin}\ \emph {et~al.}(2016)\citenamefont
  {Ashikhmin}, \citenamefont {Lai},\ and\ \citenamefont
  {Brun}}]{Ashikhmin-Lai-Brun-2016}%
  \BibitemOpen
  \bibfield  {author} {\bibinfo {author} {\bibfnamefont {A.}~\bibnamefont
  {Ashikhmin}}, \bibinfo {author} {\bibfnamefont {C.~Y.}\ \bibnamefont {Lai}},\
  and\ \bibinfo {author} {\bibfnamefont {T.~A.}\ \bibnamefont {Brun}},\
  }\bibfield  {title} {\bibinfo {title} {Correction of data and syndrome errors
  by stabilizer codes},\ }in\ \href {https://doi.org/10.1109/ISIT.2016.7541704}
  {\emph {\bibinfo {booktitle} {2016 IEEE International Symposium on
  Information Theory (ISIT)}}}\ (\bibinfo {year} {2016})\ pp.\ \bibinfo {pages}
  {2274--2278},\ \Eprint {https://arxiv.org/abs/arXiv:1602.01545}
  {arXiv:1602.01545} \BibitemShut {NoStop}%
\bibitem [{\citenamefont {Quintavalle}\ \emph {et~al.}(2021)\citenamefont
  {Quintavalle}, \citenamefont {Vasmer}, \citenamefont {Roffe},\ and\
  \citenamefont {Campbell}}]{Quintavalle-Vasmer-Roffe-Campbell-2021}%
  \BibitemOpen
  \bibfield  {author} {\bibinfo {author} {\bibfnamefont {A.~O.}\ \bibnamefont
  {Quintavalle}}, \bibinfo {author} {\bibfnamefont {M.}~\bibnamefont {Vasmer}},
  \bibinfo {author} {\bibfnamefont {J.}~\bibnamefont {Roffe}},\ and\ \bibinfo
  {author} {\bibfnamefont {E.~T.}\ \bibnamefont {Campbell}},\ }\bibfield
  {title} {\bibinfo {title} {Single-shot error correction of three-dimensional
  homological product codes},\ }\href
  {https://doi.org/10.1103/PRXQuantum.2.020340} {\bibfield  {journal} {\bibinfo
   {journal} {PRX Quantum}\ }\textbf {\bibinfo {volume} {2}},\ \bibinfo {pages}
  {020340} (\bibinfo {year} {2021})}\BibitemShut {NoStop}%
\bibitem [{\citenamefont {Breuckmann}\ and\ \citenamefont
  {Londe}(2020)}]{Breuckmann-Londe-2020}%
  \BibitemOpen
  \bibfield  {author} {\bibinfo {author} {\bibfnamefont {N.~P.}\ \bibnamefont
  {Breuckmann}}\ and\ \bibinfo {author} {\bibfnamefont {V.}~\bibnamefont
  {Londe}},\ }\bibfield  {title} {\bibinfo {title} {Single-shot decoding of
  linear rate {LDPC} quantum codes with high performance},\ }\Eprint
  {https://arxiv.org/abs/arXiv:2001.03568} {arXiv:2001.03568}  (\bibinfo {year}
  {2020}),\ \bibinfo {note} {unpublished}\BibitemShut {NoStop}%
\bibitem [{\citenamefont {Higgott}\ and\ \citenamefont
  {Breuckmann}(2023)}]{Higgott-Breuckmann-2023}%
  \BibitemOpen
  \bibfield  {author} {\bibinfo {author} {\bibfnamefont {O.}~\bibnamefont
  {Higgott}}\ and\ \bibinfo {author} {\bibfnamefont {N.~P.}\ \bibnamefont
  {Breuckmann}},\ }\bibfield  {title} {\bibinfo {title} {Improved single-shot
  decoding of higher-dimensional hypergraph-product codes},\ }\href
  {https://doi.org/10.1103/PRXQuantum.4.020332} {\bibfield  {journal} {\bibinfo
   {journal} {PRX Quantum}\ }\textbf {\bibinfo {volume} {4}},\ \bibinfo {pages}
  {020332} (\bibinfo {year} {2023})}\BibitemShut {NoStop}%
\bibitem [{\citenamefont {Lin}\ \emph {et~al.}(2024)\citenamefont {Lin},
  \citenamefont {Huang},\ and\ \citenamefont {Brown}}]{Lin-Huang-Brown-2023}%
  \BibitemOpen
  \bibfield  {author} {\bibinfo {author} {\bibfnamefont {Y.}~\bibnamefont
  {Lin}}, \bibinfo {author} {\bibfnamefont {S.}~\bibnamefont {Huang}},\ and\
  \bibinfo {author} {\bibfnamefont {K.~R.}\ \bibnamefont {Brown}},\ }\bibfield
  {title} {\bibinfo {title} {Single-shot error correction on toric codes with
  high-weight stabilizers},\ }\href
  {https://doi.org/10.1103/PhysRevA.109.052438} {\bibfield  {journal} {\bibinfo
   {journal} {Phys. Rev. A}\ }\textbf {\bibinfo {volume} {109}},\ \bibinfo
  {pages} {052438} (\bibinfo {year} {2024})},\ \Eprint
  {https://arxiv.org/abs/2310.16160} {2310.16160} \BibitemShut {NoStop}%
\bibitem [{\citenamefont {Kovalev}\ and\ \citenamefont
  {Pryadko}(2013)}]{Kovalev-Pryadko-Hyperbicycle-2013}%
  \BibitemOpen
  \bibfield  {author} {\bibinfo {author} {\bibfnamefont {A.~A.}\ \bibnamefont
  {Kovalev}}\ and\ \bibinfo {author} {\bibfnamefont {L.~P.}\ \bibnamefont
  {Pryadko}},\ }\bibfield  {title} {\bibinfo {title} {Quantum {K}ronecker
  sum-product low-density parity-check codes with finite rate},\ }\href
  {https://doi.org/10.1103/PhysRevA.88.012311} {\bibfield  {journal} {\bibinfo
  {journal} {Phys. Rev. A}\ }\textbf {\bibinfo {volume} {88}},\ \bibinfo
  {pages} {012311} (\bibinfo {year} {2013})}\BibitemShut {NoStop}%
\bibitem [{\citenamefont {Panteleev}\ and\ \citenamefont
  {Kalachev}(2021)}]{Panteleev-Kalachev-2019}%
  \BibitemOpen
  \bibfield  {author} {\bibinfo {author} {\bibfnamefont {P.}~\bibnamefont
  {Panteleev}}\ and\ \bibinfo {author} {\bibfnamefont {G.}~\bibnamefont
  {Kalachev}},\ }\bibfield  {title} {\bibinfo {title} {Degenerate quantum
  {LDPC} codes with good finite length performance},\ }\href
  {https://doi.org/10.22331/q-2021-11-22-585} {\bibfield  {journal} {\bibinfo
  {journal} {Quantum}\ }\textbf {\bibinfo {volume} {5}},\ \bibinfo {pages}
  {585} (\bibinfo {year} {2021})},\ \Eprint {https://arxiv.org/abs/1904.02703}
  {1904.02703} \BibitemShut {NoStop}%
\bibitem [{\citenamefont {Wang}\ and\ \citenamefont
  {Pryadko}(2022)}]{Wang-Pryadko-2022}%
  \BibitemOpen
  \bibfield  {author} {\bibinfo {author} {\bibfnamefont {R.}~\bibnamefont
  {Wang}}\ and\ \bibinfo {author} {\bibfnamefont {L.~P.}\ \bibnamefont
  {Pryadko}},\ }\bibfield  {title} {\bibinfo {title} {Distance bounds for
  generalized bicycle codes},\ }\href {https://doi.org/10.3390/sym14071348}
  {\bibfield  {journal} {\bibinfo  {journal} {Symmetry}\ }\textbf {\bibinfo
  {volume} {14}},\ \bibinfo {pages} {1348} (\bibinfo {year}
  {2022})}\BibitemShut {NoStop}%
\bibitem [{\citenamefont {Wang}\ \emph {et~al.}(2023)\citenamefont {Wang},
  \citenamefont {Lin},\ and\ \citenamefont {Pryadko}}]{Wang-Lin-Pryadko-2023}%
  \BibitemOpen
  \bibfield  {author} {\bibinfo {author} {\bibfnamefont {R.}~\bibnamefont
  {Wang}}, \bibinfo {author} {\bibfnamefont {H.-K.}\ \bibnamefont {Lin}},\ and\
  \bibinfo {author} {\bibfnamefont {L.~P.}\ \bibnamefont {Pryadko}},\
  }\bibfield  {title} {\bibinfo {title} {Abelian and nonabelian quantum
  two-block codes},\ }\Eprint {https://arxiv.org/abs/arXiv:2305.06890}
  {arXiv:2305.06890}  (\bibinfo {year} {2023}),\ \bibinfo {note} {to be
  publised in the proceedings of the International Symposium on Topics in
  Coding (ISTC-2023), September 4--8, 2023, Brest, France}\BibitemShut
  {NoStop}%
\bibitem [{\citenamefont {Lin}\ and\ \citenamefont
  {Pryadko}(2023)}]{Lin-Pryadko-2023}%
  \BibitemOpen
  \bibfield  {author} {\bibinfo {author} {\bibfnamefont {H.-K.}\ \bibnamefont
  {Lin}}\ and\ \bibinfo {author} {\bibfnamefont {L.~P.}\ \bibnamefont
  {Pryadko}},\ }\bibfield  {title} {\bibinfo {title} {Quantum two-block group
  algebra codes},\ }\Eprint {https://arxiv.org/abs/arXiv:2306.16400}
  {arXiv:2306.16400}  (\bibinfo {year} {2023}),\ \bibinfo {note}
  {unpublished}\BibitemShut {NoStop}%
\bibitem [{\citenamefont {Bravyi}\ \emph {et~al.}(2023)\citenamefont {Bravyi},
  \citenamefont {Cross}, \citenamefont {Gambetta}, \citenamefont {Maslov},
  \citenamefont {Rall},\ and\ \citenamefont {Yoder}}]{Bravyi-etal-Yoder-2023}%
  \BibitemOpen
  \bibfield  {author} {\bibinfo {author} {\bibfnamefont {S.}~\bibnamefont
  {Bravyi}}, \bibinfo {author} {\bibfnamefont {A.~W.}\ \bibnamefont {Cross}},
  \bibinfo {author} {\bibfnamefont {J.~M.}\ \bibnamefont {Gambetta}}, \bibinfo
  {author} {\bibfnamefont {D.}~\bibnamefont {Maslov}}, \bibinfo {author}
  {\bibfnamefont {P.}~\bibnamefont {Rall}},\ and\ \bibinfo {author}
  {\bibfnamefont {T.~J.}\ \bibnamefont {Yoder}},\ }\bibfield  {title} {\bibinfo
  {title} {High-threshold and low-overhead fault-tolerant quantum memory},\
  }\Eprint {https://arxiv.org/abs/2308.07915} {arXiv:2308.07915 [quant-ph]}
  (\bibinfo {year} {2023}),\ \bibinfo {note} {unpublished}\BibitemShut
  {NoStop}%
\bibitem [{\citenamefont {Wang}\ \emph {et~al.}(2011)\citenamefont {Wang},
  \citenamefont {Fowler},\ and\ \citenamefont
  {Hollenberg}}]{Wang-Fowler-Hollenberg-2011}%
  \BibitemOpen
  \bibfield  {author} {\bibinfo {author} {\bibfnamefont {D.~S.}\ \bibnamefont
  {Wang}}, \bibinfo {author} {\bibfnamefont {A.~G.}\ \bibnamefont {Fowler}},\
  and\ \bibinfo {author} {\bibfnamefont {L.~C.~L.}\ \bibnamefont
  {Hollenberg}},\ }\bibfield  {title} {\bibinfo {title} {Surface code quantum
  computing with error rates over {$1\%$}},\ }\href
  {https://doi.org/10.1103/PhysRevA.83.020302} {\bibfield  {journal} {\bibinfo
  {journal} {Phys. Rev. A}\ }\textbf {\bibinfo {volume} {83}},\ \bibinfo
  {pages} {020302} (\bibinfo {year} {2011})}\BibitemShut {NoStop}%
\bibitem [{\citenamefont {Roffe}\ \emph {et~al.}(2020)\citenamefont {Roffe},
  \citenamefont {White}, \citenamefont {Burton},\ and\ \citenamefont
  {Campbell}}]{Roffe-White-Burton-Campbell-2020}%
  \BibitemOpen
  \bibfield  {author} {\bibinfo {author} {\bibfnamefont {J.}~\bibnamefont
  {Roffe}}, \bibinfo {author} {\bibfnamefont {D.~R.}\ \bibnamefont {White}},
  \bibinfo {author} {\bibfnamefont {S.}~\bibnamefont {Burton}},\ and\ \bibinfo
  {author} {\bibfnamefont {E.}~\bibnamefont {Campbell}},\ }\bibfield  {title}
  {\bibinfo {title} {Decoding across the quantum low-density parity-check code
  landscape},\ }\href {https://doi.org/10.1103/PhysRevResearch.2.043423}
  {\bibfield  {journal} {\bibinfo  {journal} {Phys. Rev. Res.}\ }\textbf
  {\bibinfo {volume} {2}},\ \bibinfo {pages} {043423} (\bibinfo {year}
  {2020})},\ \Eprint {https://arxiv.org/abs/arXiv:2005.07016}
  {arXiv:2005.07016} \BibitemShut {NoStop}%
\bibitem [{\citenamefont {Tan}\ \emph {et~al.}(2023)\citenamefont {Tan},
  \citenamefont {Zhang}, \citenamefont {Chao}, \citenamefont {Shi},\ and\
  \citenamefont {Chen}}]{Tan-Zhang-Chao-Shi-Chen-2023}%
  \BibitemOpen
  \bibfield  {author} {\bibinfo {author} {\bibfnamefont {X.}~\bibnamefont
  {Tan}}, \bibinfo {author} {\bibfnamefont {F.}~\bibnamefont {Zhang}}, \bibinfo
  {author} {\bibfnamefont {R.}~\bibnamefont {Chao}}, \bibinfo {author}
  {\bibfnamefont {Y.}~\bibnamefont {Shi}},\ and\ \bibinfo {author}
  {\bibfnamefont {J.}~\bibnamefont {Chen}},\ }\bibfield  {title} {\bibinfo
  {title} {Scalable surface-code decoders with parallelization in time},\
  }\href {https://doi.org/10.1103/PRXQuantum.4.040344} {\bibfield  {journal}
  {\bibinfo  {journal} {PRX Quantum}\ }\textbf {\bibinfo {volume} {4}},\
  \bibinfo {pages} {040344} (\bibinfo {year} {2023})}\BibitemShut {NoStop}%
\bibitem [{\citenamefont {Kalachev}\ and\ \citenamefont
  {Panteleev}(2020)}]{Kalachev-Panteleev-2020}%
  \BibitemOpen
  \bibfield  {author} {\bibinfo {author} {\bibfnamefont {G.~V.}\ \bibnamefont
  {Kalachev}}\ and\ \bibinfo {author} {\bibfnamefont {P.~A.}\ \bibnamefont
  {Panteleev}},\ }\bibfield  {title} {\bibinfo {title} {On the minimum distance
  in one class of quantum {LDPC} codes},\ }\href
  {http://mi.mathnet.ru/eng/ista284} {\bibfield  {journal} {\bibinfo  {journal}
  {Intelligent systems. Theory and applications}\ }\textbf {\bibinfo {volume}
  {24}},\ \bibinfo {pages} {87–117} (\bibinfo {year} {2020})},\ \bibinfo
  {note} {[In Russian]}\BibitemShut {NoStop}%
\bibitem [{\citenamefont {MacWilliams}\ and\ \citenamefont
  {Sloane}(1981)}]{MS-book}%
  \BibitemOpen
  \bibfield  {author} {\bibinfo {author} {\bibfnamefont {F.~J.}\ \bibnamefont
  {MacWilliams}}\ and\ \bibinfo {author} {\bibfnamefont {N.~J.~A.}\
  \bibnamefont {Sloane}},\ }\href@noop {} {\emph {\bibinfo {title} {The Theory
  of Error-Correcting Codes}}}\ (\bibinfo  {publisher} {North-Holland},\
  \bibinfo {address} {Amsterdam},\ \bibinfo {year} {1981})\BibitemShut
  {NoStop}%
\bibitem [{\citenamefont {Kitaev}(2003)}]{kitaev-anyons}%
  \BibitemOpen
  \bibfield  {author} {\bibinfo {author} {\bibfnamefont {A.~Y.}\ \bibnamefont
  {Kitaev}},\ }\bibfield  {title} {\bibinfo {title} {Fault-tolerant quantum
  computation by anyons},\ }\href {http://arxiv.org/abs/quant-ph/9707021}
  {\bibfield  {journal} {\bibinfo  {journal} {Ann. Phys.}\ }\textbf {\bibinfo
  {volume} {303}},\ \bibinfo {pages} {2} (\bibinfo {year} {2003})}\BibitemShut
  {NoStop}%
\bibitem [{\citenamefont {Bravyi}\ and\ \citenamefont
  {Kitaev}(1998)}]{Bravyi-Kitaev-1998}%
  \BibitemOpen
  \bibfield  {author} {\bibinfo {author} {\bibfnamefont {S.~B.}\ \bibnamefont
  {Bravyi}}\ and\ \bibinfo {author} {\bibfnamefont {A.~Y.}\ \bibnamefont
  {Kitaev}},\ }\bibfield  {title} {\bibinfo {title} {Quantum codes on a lattice
  with boundary},\ }\Eprint {https://arxiv.org/abs/quant-ph/9811052}
  {quant-ph/9811052}  (\bibinfo {year} {1998}),\ \bibinfo {note}
  {unpublished}\BibitemShut {NoStop}%
\bibitem [{\citenamefont {Freedman}\ and\ \citenamefont
  {Meyer}(2001)}]{Freedman-Meyer-1998}%
  \BibitemOpen
  \bibfield  {author} {\bibinfo {author} {\bibfnamefont {M.~H.}\ \bibnamefont
  {Freedman}}\ and\ \bibinfo {author} {\bibfnamefont {D.~A.}\ \bibnamefont
  {Meyer}},\ }\bibfield  {title} {\bibinfo {title} {Projective plane and planar
  quantum codes},\ }\href {https://doi.org/10.1007/s102080010013} {\bibfield
  {journal} {\bibinfo  {journal} {Foundations of Computational Mathematics}\
  }\textbf {\bibinfo {volume} {1}},\ \bibinfo {pages} {325} (\bibinfo {year}
  {2001})},\ \Eprint {https://arxiv.org/abs/quant-ph/9810055}
  {quant-ph/9810055} \BibitemShut {NoStop}%
\bibitem [{\citenamefont {Bombin}\ and\ \citenamefont
  {Martin-Delgado}(2007)}]{Bombin-MartinDelgado-2007}%
  \BibitemOpen
  \bibfield  {author} {\bibinfo {author} {\bibfnamefont {H.}~\bibnamefont
  {Bombin}}\ and\ \bibinfo {author} {\bibfnamefont {M.~A.}\ \bibnamefont
  {Martin-Delgado}},\ }\bibfield  {title} {\bibinfo {title} {Homological error
  correction: {C}lassical and quantum codes},\ }\href
  {https://doi.org/10.1063/1.2731356} {\bibfield  {journal} {\bibinfo
  {journal} {Journal of Mathematical Physics}\ }\textbf {\bibinfo {volume}
  {48}},\ \bibinfo {eid} {052105} (\bibinfo {year} {2007})}\BibitemShut
  {NoStop}%
\bibitem [{\citenamefont {Castelnovo}\ and\ \citenamefont
  {Chamon}(2008)}]{Castelnovo-Chamon-2008}%
  \BibitemOpen
  \bibfield  {author} {\bibinfo {author} {\bibfnamefont {C.}~\bibnamefont
  {Castelnovo}}\ and\ \bibinfo {author} {\bibfnamefont {C.}~\bibnamefont
  {Chamon}},\ }\bibfield  {title} {\bibinfo {title} {Topological order in a
  three-dimensional toric code at finite temperature},\ }\href
  {https://doi.org/10.1103/PhysRevB.78.155120} {\bibfield  {journal} {\bibinfo
  {journal} {Phys. Rev. B}\ }\textbf {\bibinfo {volume} {78}},\ \bibinfo
  {pages} {155120} (\bibinfo {year} {2008})}\BibitemShut {NoStop}%
\bibitem [{\citenamefont {Maz{\'{a}\v{c}}}\ and\ \citenamefont
  {Hamma}(2012)}]{Mazac-Hamma-2012}%
  \BibitemOpen
  \bibfield  {author} {\bibinfo {author} {\bibfnamefont {D.}~\bibnamefont
  {Maz{\'{a}\v{c}}}}\ and\ \bibinfo {author} {\bibfnamefont {A.}~\bibnamefont
  {Hamma}},\ }\bibfield  {title} {\bibinfo {title} {Topological order,
  entanglement, and quantum memory at finite temperature},\ }\href
  {https://doi.org/10.1016/j.aop.2012.05.004} {\bibfield  {journal} {\bibinfo
  {journal} {Annals of Physics}\ }\textbf {\bibinfo {volume} {327}},\ \bibinfo
  {pages} {2096 } (\bibinfo {year} {2012})}\BibitemShut {NoStop}%
\bibitem [{\citenamefont {Bombin}\ \emph {et~al.}(2013)\citenamefont {Bombin},
  \citenamefont {Chhajlany}, \citenamefont {Horodecki},\ and\ \citenamefont
  {Martin-Delgado}}]{Bombin-Chhajlany-Horodecki-MartinDelgado-2013}%
  \BibitemOpen
  \bibfield  {author} {\bibinfo {author} {\bibfnamefont {H.}~\bibnamefont
  {Bombin}}, \bibinfo {author} {\bibfnamefont {R.~W.}\ \bibnamefont
  {Chhajlany}}, \bibinfo {author} {\bibfnamefont {M.}~\bibnamefont
  {Horodecki}},\ and\ \bibinfo {author} {\bibfnamefont {M.~A.}\ \bibnamefont
  {Martin-Delgado}},\ }\bibfield  {title} {\bibinfo {title} {Self-correcting
  quantum computers},\ }\href {http://stacks.iop.org/1367-2630/15/i=5/a=055023}
  {\bibfield  {journal} {\bibinfo  {journal} {New Journal of Physics}\ }\textbf
  {\bibinfo {volume} {15}},\ \bibinfo {pages} {055023} (\bibinfo {year}
  {2013})}\BibitemShut {NoStop}%
\bibitem [{\citenamefont {Tillich}\ and\ \citenamefont
  {Z{\'e}mor}(2009)}]{Tillich-Zemor-2009}%
  \BibitemOpen
  \bibfield  {author} {\bibinfo {author} {\bibfnamefont {J.-P.}\ \bibnamefont
  {Tillich}}\ and\ \bibinfo {author} {\bibfnamefont {G.}~\bibnamefont
  {Z{\'e}mor}},\ }\bibfield  {title} {\bibinfo {title} {Quantum {LDPC} codes
  with positive rate and minimum distance proportional to {$\sqrt{n}$}},\ }in\
  \href {https://doi.org/10.1109/ISIT.2009.5205648} {\emph {\bibinfo
  {booktitle} {Proc. IEEE Int. Symp. Inf. Theory (ISIT)}}}\ (\bibinfo {year}
  {2009})\ pp.\ \bibinfo {pages} {799--803}\BibitemShut {NoStop}%
\bibitem [{\citenamefont {Zeng}\ and\ \citenamefont
  {Pryadko}(2019)}]{Zeng-Pryadko-2018}%
  \BibitemOpen
  \bibfield  {author} {\bibinfo {author} {\bibfnamefont {W.}~\bibnamefont
  {Zeng}}\ and\ \bibinfo {author} {\bibfnamefont {L.~P.}\ \bibnamefont
  {Pryadko}},\ }\bibfield  {title} {\bibinfo {title} {Higher-dimensional
  quantum hypergraph-product codes with finite rates},\ }\href
  {https://doi.org/10.1103/PhysRevLett.122.230501} {\bibfield  {journal}
  {\bibinfo  {journal} {Phys. Rev. Lett.}\ }\textbf {\bibinfo {volume} {122}},\
  \bibinfo {pages} {230501} (\bibinfo {year} {2019})},\ \Eprint
  {https://arxiv.org/abs/1810.01519} {1810.01519} \BibitemShut {NoStop}%
\bibitem [{\citenamefont {Zeng}\ and\ \citenamefont
  {Pryadko}(2020)}]{Zeng-Pryadko-hprod-2020}%
  \BibitemOpen
  \bibfield  {author} {\bibinfo {author} {\bibfnamefont {W.}~\bibnamefont
  {Zeng}}\ and\ \bibinfo {author} {\bibfnamefont {L.~P.}\ \bibnamefont
  {Pryadko}},\ }\bibfield  {title} {\bibinfo {title} {Minimal distances for
  certain quantum product codes and tensor products of chain complexes},\
  }\href {https://doi.org/10.1103/PhysRevA.102.062402} {\bibfield  {journal}
  {\bibinfo  {journal} {Phys. Rev. A}\ }\textbf {\bibinfo {volume} {102}},\
  \bibinfo {pages} {062402} (\bibinfo {year} {2020})},\ \Eprint
  {https://arxiv.org/abs/arXiv:2007.12152} {arXiv:2007.12152} \BibitemShut
  {NoStop}%
\bibitem [{\citenamefont {Raussendorf}\ and\ \citenamefont
  {Harrington}(2007)}]{Raussendorf-Harrington-2007}%
  \BibitemOpen
  \bibfield  {author} {\bibinfo {author} {\bibfnamefont {R.}~\bibnamefont
  {Raussendorf}}\ and\ \bibinfo {author} {\bibfnamefont {J.}~\bibnamefont
  {Harrington}},\ }\bibfield  {title} {\bibinfo {title} {Fault-tolerant quantum
  computation with high threshold in two dimensions},\ }\href
  {http://link.aps.org/abstract/PRL/v98/e190504} {\bibfield  {journal}
  {\bibinfo  {journal} {Phys. Rev. Lett.}\ }\textbf {\bibinfo {volume} {98}},\
  \bibinfo {pages} {190504} (\bibinfo {year} {2007})}\BibitemShut {NoStop}%
\bibitem [{\citenamefont {Huang}\ \emph {et~al.}(2020)\citenamefont {Huang},
  \citenamefont {Newman},\ and\ \citenamefont
  {Brown}}]{Huang-Newman-Brown-2020}%
  \BibitemOpen
  \bibfield  {author} {\bibinfo {author} {\bibfnamefont {S.}~\bibnamefont
  {Huang}}, \bibinfo {author} {\bibfnamefont {M.}~\bibnamefont {Newman}},\ and\
  \bibinfo {author} {\bibfnamefont {K.~R.}\ \bibnamefont {Brown}},\ }\bibfield
  {title} {\bibinfo {title} {Fault-tolerant weighted union-find decoding on the
  toric code},\ }\href {https://doi.org/10.1103/PhysRevA.102.012419} {\bibfield
   {journal} {\bibinfo  {journal} {Phys. Rev. A}\ }\textbf {\bibinfo {volume}
  {102}},\ \bibinfo {pages} {012419} (\bibinfo {year} {2020})}\BibitemShut
  {NoStop}%
\bibitem [{\citenamefont {Gidney}(2021)}]{Gidney-2021-stim}%
  \BibitemOpen
  \bibfield  {author} {\bibinfo {author} {\bibfnamefont {C.}~\bibnamefont
  {Gidney}},\ }\bibfield  {title} {\bibinfo {title} {Stim: a fast stabilizer
  circuit simulator},\ }\href {https://doi.org/10.22331/q-2021-07-06-497}
  {\bibfield  {journal} {\bibinfo  {journal} {{Quantum}}\ }\textbf {\bibinfo
  {volume} {5}},\ \bibinfo {pages} {497} (\bibinfo {year} {2021})}\BibitemShut
  {NoStop}%
\bibitem [{\citenamefont {Pryadko}(2020)}]{Pryadko-2020}%
  \BibitemOpen
  \bibfield  {author} {\bibinfo {author} {\bibfnamefont {L.~P.}\ \bibnamefont
  {Pryadko}},\ }\bibfield  {title} {\bibinfo {title} {On maximum-likelihood
  decoding with circuit-level errors},\ }\href
  {https://doi.org/10.22331/q-2020-08-06-304} {\bibfield  {journal} {\bibinfo
  {journal} {Quantum}\ }\textbf {\bibinfo {volume} {4}},\ \bibinfo {pages}
  {304} (\bibinfo {year} {2020})},\ \Eprint
  {https://arxiv.org/abs/arXiv:1909.06732} {arXiv:1909.06732} \BibitemShut
  {NoStop}%
\bibitem [{\citenamefont {{Bacon}}\ \emph {et~al.}(2017)\citenamefont
  {{Bacon}}, \citenamefont {{Flammia}}, \citenamefont {{Harrow}},\ and\
  \citenamefont {{Shi}}}]{Bacon-Flammia-Harrow-Shi-2017}%
  \BibitemOpen
  \bibfield  {author} {\bibinfo {author} {\bibfnamefont {D.}~\bibnamefont
  {{Bacon}}}, \bibinfo {author} {\bibfnamefont {S.~T.}\ \bibnamefont
  {{Flammia}}}, \bibinfo {author} {\bibfnamefont {A.~W.}\ \bibnamefont
  {{Harrow}}},\ and\ \bibinfo {author} {\bibfnamefont {J.}~\bibnamefont
  {{Shi}}},\ }\bibfield  {title} {\bibinfo {title} {Sparse quantum codes from
  quantum circuits},\ }\href {https://doi.org/10.1109/TIT.2017.2663199}
  {\bibfield  {journal} {\bibinfo  {journal} {IEEE Transactions on Information
  Theory}\ }\textbf {\bibinfo {volume} {63}},\ \bibinfo {pages} {2464}
  (\bibinfo {year} {2017})}\BibitemShut {NoStop}%
\bibitem [{\citenamefont {Tomita}\ and\ \citenamefont
  {Svore}(2014)}]{Tomita-Svore-2014}%
  \BibitemOpen
  \bibfield  {author} {\bibinfo {author} {\bibfnamefont {Y.}~\bibnamefont
  {Tomita}}\ and\ \bibinfo {author} {\bibfnamefont {K.~M.}\ \bibnamefont
  {Svore}},\ }\bibfield  {title} {\bibinfo {title} {Low-distance surface codes
  under realistic quantum noise},\ }\href
  {https://doi.org/10.1103/PhysRevA.90.062320} {\bibfield  {journal} {\bibinfo
  {journal} {Phys. Rev. A}\ }\textbf {\bibinfo {volume} {90}},\ \bibinfo
  {pages} {062320} (\bibinfo {year} {2014})},\ \Eprint
  {https://arxiv.org/abs/1404.3747} {1404.3747} \BibitemShut {NoStop}%
\bibitem [{\citenamefont {Kovalev}\ and\ \citenamefont
  {Pryadko}(2015)}]{Kovalev-Pryadko-SG-2015}%
  \BibitemOpen
  \bibfield  {author} {\bibinfo {author} {\bibfnamefont {A.~A.}\ \bibnamefont
  {Kovalev}}\ and\ \bibinfo {author} {\bibfnamefont {L.~P.}\ \bibnamefont
  {Pryadko}},\ }\bibfield  {title} {\bibinfo {title} {Spin glass reflection of
  the decoding transition for quantum error-correcting codes},\ }\href@noop {}
  {\bibfield  {journal} {\bibinfo  {journal} {Quantum Inf. {\&} Comp.}\
  }\textbf {\bibinfo {volume} {15}},\ \bibinfo {pages} {0825} (\bibinfo {year}
  {2015})},\ \Eprint {https://arxiv.org/abs/arXiv:1311.7688} {arXiv:1311.7688}
  \BibitemShut {NoStop}%
\bibitem [{\citenamefont {Iyer}\ and\ \citenamefont
  {Poulin}(2015)}]{Iyer-Poulin-2013}%
  \BibitemOpen
  \bibfield  {author} {\bibinfo {author} {\bibfnamefont {P.}~\bibnamefont
  {Iyer}}\ and\ \bibinfo {author} {\bibfnamefont {D.}~\bibnamefont {Poulin}},\
  }\bibfield  {title} {\bibinfo {title} {Hardness of decoding quantum
  stabilizer codes},\ }\href {https://doi.org/10.1109/TIT.2015.2422294}
  {\bibfield  {journal} {\bibinfo  {journal} {{IEEE} Transactions on
  Information Theory}\ }\textbf {\bibinfo {volume} {61}},\ \bibinfo {pages}
  {5209} (\bibinfo {year} {2015})},\ \Eprint
  {https://arxiv.org/abs/arXiv:1310.3235} {arXiv:1310.3235} \BibitemShut
  {NoStop}%
\bibitem [{\citenamefont {Andriyanova}\ \emph {et~al.}(2012)\citenamefont
  {Andriyanova}, \citenamefont {Maurice},\ and\ \citenamefont
  {Tillich}}]{Andriyanova-Maurice-Tillich-2012}%
  \BibitemOpen
  \bibfield  {author} {\bibinfo {author} {\bibfnamefont {I.}~\bibnamefont
  {Andriyanova}}, \bibinfo {author} {\bibfnamefont {D.}~\bibnamefont
  {Maurice}},\ and\ \bibinfo {author} {\bibfnamefont {J.-P.}\ \bibnamefont
  {Tillich}},\ }\bibfield  {title} {\bibinfo {title} {New constructions of
  {CSS} codes obtained by moving to higher alphabets},\ }\Eprint
  {https://arxiv.org/abs/arXiv:1202.3338} {arXiv:1202.3338}  (\bibinfo {year}
  {2012}),\ \bibinfo {note} {unpublished}\BibitemShut {NoStop}%
\bibitem [{\citenamefont {{Xie}}\ and\ \citenamefont
  {{Yuan}}(2016)}]{Xie-Yuan-2016}%
  \BibitemOpen
  \bibfield  {author} {\bibinfo {author} {\bibfnamefont {Y.}~\bibnamefont
  {{Xie}}}\ and\ \bibinfo {author} {\bibfnamefont {J.}~\bibnamefont {{Yuan}}},\
  }\bibfield  {title} {\bibinfo {title} {Reliable quantum ldpc codes over
  gf(4)},\ }in\ \href {https://doi.org/10.1109/GLOCOMW.2016.7849021} {\emph
  {\bibinfo {booktitle} {2016 IEEE Globecom Workshops (GC Wkshps)}}}\ (\bibinfo
  {year} {2016})\ pp.\ \bibinfo {pages} {1--5}\BibitemShut {NoStop}%
\bibitem [{\citenamefont {{Kuo}}\ and\ \citenamefont
  {{Lai}}(2020)}]{Kuo-Lai-2020}%
  \BibitemOpen
  \bibfield  {author} {\bibinfo {author} {\bibfnamefont {K.~Y.}\ \bibnamefont
  {{Kuo}}}\ and\ \bibinfo {author} {\bibfnamefont {C.~Y.}\ \bibnamefont
  {{Lai}}},\ }\bibfield  {title} {\bibinfo {title} {Refined belief propagation
  decoding of sparse-graph quantum codes},\ }\href
  {https://doi.org/10.1109/JSAIT.2020.3011758} {\bibfield  {journal} {\bibinfo
  {journal} {IEEE Journal on Selected Areas in Information Theory}\ }\textbf
  {\bibinfo {volume} {1}},\ \bibinfo {pages} {487} (\bibinfo {year}
  {2020})}\BibitemShut {NoStop}%
\bibitem [{\citenamefont {Richardson}\ \emph {et~al.}(2001)\citenamefont
  {Richardson}, \citenamefont {Shokrollahi},\ and\ \citenamefont
  {Urbanke}}]{Richardson-Shokrollahi-Amin-Urbanke-2001}%
  \BibitemOpen
  \bibfield  {author} {\bibinfo {author} {\bibfnamefont {T.~J.}\ \bibnamefont
  {Richardson}}, \bibinfo {author} {\bibfnamefont {M.~A.}\ \bibnamefont
  {Shokrollahi}},\ and\ \bibinfo {author} {\bibfnamefont {R.~L.}\ \bibnamefont
  {Urbanke}},\ }\bibfield  {title} {\bibinfo {title} {Design of
  capacity-approaching irregular low-density parity-check codes},\ }\href
  {https://doi.org/10.1109/18.910578} {\bibfield  {journal} {\bibinfo
  {journal} {Information Theory, IEEE Transactions on}\ }\textbf {\bibinfo
  {volume} {47}},\ \bibinfo {pages} {619} (\bibinfo {year} {2001})}\BibitemShut
  {NoStop}%
\bibitem [{\citenamefont {Chung}\ \emph {et~al.}(2001)\citenamefont {Chung},
  \citenamefont {Forney~Jr}, \citenamefont {Richardson},\ and\ \citenamefont
  {Urbanke}}]{Chung-Forney-Richardson-Urbanke-2001}%
  \BibitemOpen
  \bibfield  {author} {\bibinfo {author} {\bibfnamefont {S.-Y.}\ \bibnamefont
  {Chung}}, \bibinfo {author} {\bibfnamefont {G.~D.}\ \bibnamefont
  {Forney~Jr}}, \bibinfo {author} {\bibfnamefont {T.~J.}\ \bibnamefont
  {Richardson}},\ and\ \bibinfo {author} {\bibfnamefont {R.}~\bibnamefont
  {Urbanke}},\ }\bibfield  {title} {\bibinfo {title} {On the design of
  low-density parity-check codes within 0.0045 d{B} of the {S}hannon limit},\
  }\href@noop {} {\bibfield  {journal} {\bibinfo  {journal} {Communications
  Letters, IEEE}\ }\textbf {\bibinfo {volume} {5}},\ \bibinfo {pages} {58}
  (\bibinfo {year} {2001})}\BibitemShut {NoStop}%
\bibitem [{\citenamefont {Poulin}\ and\ \citenamefont
  {Chung}(2008)}]{Poulin-Chung-2008}%
  \BibitemOpen
  \bibfield  {author} {\bibinfo {author} {\bibfnamefont {D.}~\bibnamefont
  {Poulin}}\ and\ \bibinfo {author} {\bibfnamefont {Y.}~\bibnamefont {Chung}},\
  }\bibfield  {title} {\bibinfo {title} {On the iterative decoding of sparse
  quantum codes},\ }\href@noop {} {\bibfield  {journal} {\bibinfo  {journal}
  {Quant. Info. and Comp.}\ }\textbf {\bibinfo {volume} {8}},\ \bibinfo {pages}
  {987} (\bibinfo {year} {2008})},\ \Eprint
  {https://arxiv.org/abs/arXiv:0801.1241} {arXiv:0801.1241} \BibitemShut
  {NoStop}%
\bibitem [{\citenamefont {Fossorier}\ and\ \citenamefont
  {Lin}(1995)}]{Fossorier-Lin-1995}%
  \BibitemOpen
  \bibfield  {author} {\bibinfo {author} {\bibfnamefont {M.~P.~C.}\
  \bibnamefont {Fossorier}}\ and\ \bibinfo {author} {\bibfnamefont
  {S.}~\bibnamefont {Lin}},\ }\bibfield  {title} {\bibinfo {title}
  {Soft-decision decoding of linear block codes based on ordered statistics},\
  }\href {https://doi.org/10.1109/18.412683} {\bibfield  {journal} {\bibinfo
  {journal} {IEEE Transactions on Information Theory}\ }\textbf {\bibinfo
  {volume} {41}},\ \bibinfo {pages} {1379} (\bibinfo {year}
  {1995})}\BibitemShut {NoStop}%
\bibitem [{\citenamefont {Fossorier}(2001)}]{Fossorier-2001}%
  \BibitemOpen
  \bibfield  {author} {\bibinfo {author} {\bibfnamefont {M.~P.~C.}\
  \bibnamefont {Fossorier}},\ }\bibfield  {title} {\bibinfo {title} {Iterative
  reliability-based decoding of low-density parity check codes},\ }\href
  {https://doi.org/10.1109/49.924874} {\bibfield  {journal} {\bibinfo
  {journal} {{IEEE} Journal on Selected Areas in Communications}\ }\textbf
  {\bibinfo {volume} {19}},\ \bibinfo {pages} {908} (\bibinfo {year}
  {2001})}\BibitemShut {NoStop}%
\bibitem [{\citenamefont {Pryadko}(2025)}]{Pryadko-2025-vecdec}%
  \BibitemOpen
  \bibfield  {author} {\bibinfo {author} {\bibfnamefont {L.~P.}\ \bibnamefont
  {Pryadko}},\ }\href@noop {} {\bibinfo {title} {vecdec --- vectorized decoder
  and {LER} estimator}},\ \bibinfo {howpublished}
  {\url{https://github.com/QEC-pages/vecdec}} (\bibinfo {year}
  {2025})\BibitemShut {NoStop}%
\bibitem [{\citenamefont {Prange}(1962)}]{Prange-1962}%
  \BibitemOpen
  \bibfield  {author} {\bibinfo {author} {\bibfnamefont {E.}~\bibnamefont
  {Prange}},\ }\bibfield  {title} {\bibinfo {title} {The use of information
  sets in decoding cyclic codes},\ }\href
  {https://doi.org/10.1109/TIT.1962.1057777} {\bibfield  {journal} {\bibinfo
  {journal} {Information Theory, IRE Transactions on}\ }\textbf {\bibinfo
  {volume} {8}},\ \bibinfo {pages} {5} (\bibinfo {year} {1962})}\BibitemShut
  {NoStop}%
\bibitem [{\citenamefont {Leon}(1988)}]{Leon-1988}%
  \BibitemOpen
  \bibfield  {author} {\bibinfo {author} {\bibfnamefont {J.~S.}\ \bibnamefont
  {Leon}},\ }\bibfield  {title} {\bibinfo {title} {A probabilistic algorithm
  for computing minimum weights of large error-correcting codes},\ }\href
  {https://doi.org/10.1109/18.21270} {\bibfield  {journal} {\bibinfo  {journal}
  {IEEE Trans. Info. Theory}\ }\textbf {\bibinfo {volume} {34}},\ \bibinfo
  {pages} {1354 } (\bibinfo {year} {1988})}\BibitemShut {NoStop}%
\bibitem [{\citenamefont {Kruk}(1989)}]{Kruk-1989}%
  \BibitemOpen
  \bibfield  {author} {\bibinfo {author} {\bibfnamefont {E.~A.}\ \bibnamefont
  {Kruk}},\ }\bibfield  {title} {\bibinfo {title} {Decoding complexity bound
  for linear block codes},\ }\href {http://mi.mathnet.ru/eng/ppi665} {\bibfield
   {journal} {\bibinfo  {journal} {Probl. Peredachi Inf.}\ }\textbf {\bibinfo
  {volume} {25}},\ \bibinfo {pages} {103} (\bibinfo {year} {1989})},\ \bibinfo
  {note} {(In Russian)}\BibitemShut {NoStop}%
\bibitem [{\citenamefont {Coffey}\ and\ \citenamefont
  {Goodman}(1990)}]{Coffey-Goodman-1990}%
  \BibitemOpen
  \bibfield  {author} {\bibinfo {author} {\bibfnamefont {J.~T.}\ \bibnamefont
  {Coffey}}\ and\ \bibinfo {author} {\bibfnamefont {R.~M.}\ \bibnamefont
  {Goodman}},\ }\bibfield  {title} {\bibinfo {title} {The complexity of
  information set decoding},\ }\href {https://doi.org/10.1109/18.57202}
  {\bibfield  {journal} {\bibinfo  {journal} {IEEE Trans. Info. Theory}\
  }\textbf {\bibinfo {volume} {36}},\ \bibinfo {pages} {1031 } (\bibinfo {year}
  {1990})}\BibitemShut {NoStop}%
\bibitem [{\citenamefont {Dumer}\ \emph {et~al.}(2017)\citenamefont {Dumer},
  \citenamefont {Kovalev},\ and\ \citenamefont
  {Pryadko}}]{Dumer-Kovalev-Pryadko-IEEE-2017}%
  \BibitemOpen
  \bibfield  {author} {\bibinfo {author} {\bibfnamefont {I.}~\bibnamefont
  {Dumer}}, \bibinfo {author} {\bibfnamefont {A.~A.}\ \bibnamefont {Kovalev}},\
  and\ \bibinfo {author} {\bibfnamefont {L.~P.}\ \bibnamefont {Pryadko}},\
  }\bibfield  {title} {\bibinfo {title} {Distance verification for classical
  and quantum {LDPC} codes},\ }\href {https://doi.org/10.1109/TIT.2017.2690381}
  {\bibfield  {journal} {\bibinfo  {journal} {IEEE Trans. Inf. Th.}\ }\textbf
  {\bibinfo {volume} {63}},\ \bibinfo {pages} {4675} (\bibinfo {year}
  {2017})}\BibitemShut {NoStop}%
\bibitem [{\citenamefont {Roffe}(2022)}]{Roffe_LDPC_Python_tools_2022}%
  \BibitemOpen
  \bibfield  {author} {\bibinfo {author} {\bibfnamefont {J.}~\bibnamefont
  {Roffe}},\ }\href {https://pypi.org/project/ldpc/} {\bibinfo {title} {{LDPC:
  Python tools for low density parity check codes}}} (\bibinfo {year}
  {2022})\BibitemShut {NoStop}%
\bibitem [{\citenamefont {Grospellier}\ \emph {et~al.}(2020)\citenamefont
  {Grospellier}, \citenamefont {Grou{\`e}s}, \citenamefont {Krishna},\ and\
  \citenamefont {Leverrier}}]{Grospellier-Groues-Krishna-Leverrier-2020}%
  \BibitemOpen
  \bibfield  {author} {\bibinfo {author} {\bibfnamefont {A.}~\bibnamefont
  {Grospellier}}, \bibinfo {author} {\bibfnamefont {L.}~\bibnamefont
  {Grou{\`e}s}}, \bibinfo {author} {\bibfnamefont {A.}~\bibnamefont
  {Krishna}},\ and\ \bibinfo {author} {\bibfnamefont {A.}~\bibnamefont
  {Leverrier}},\ }\bibfield  {title} {\bibinfo {title} {Combining hard and soft
  decoders for hypergraph product codes},\ }\Eprint
  {https://arxiv.org/abs/arXiv:2004.11199} {arXiv:2004.11199}  (\bibinfo {year}
  {2020}),\ \bibinfo {note} {unpublished}\BibitemShut {NoStop}%
\bibitem [{\citenamefont {Pryadko}\ \emph {et~al.}(2022)\citenamefont
  {Pryadko}, \citenamefont {Shabashov},\ and\ \citenamefont
  {Kozin}}]{Pryadko-Shabashov-Kozin-QDistRnd-2021}%
  \BibitemOpen
  \bibfield  {author} {\bibinfo {author} {\bibfnamefont {L.~P.}\ \bibnamefont
  {Pryadko}}, \bibinfo {author} {\bibfnamefont {V.~A.}\ \bibnamefont
  {Shabashov}},\ and\ \bibinfo {author} {\bibfnamefont {V.~K.}\ \bibnamefont
  {Kozin}},\ }\bibfield  {title} {\bibinfo {title} {{QDistRnd}: A {GAP} package
  for computing the distance of quantum error-correcting codes},\ }\href
  {https://doi.org/10.21105/joss.04120} {\bibfield  {journal} {\bibinfo
  {journal} {Journal of Open Source Software}\ }\textbf {\bibinfo {volume}
  {7}},\ \bibinfo {pages} {4120} (\bibinfo {year} {2022})}\BibitemShut
  {NoStop}%
\bibitem [{\citenamefont {Pryadko}\ and\ \citenamefont
  {Zeng}(2024)}]{Pryadko-Zeng-dist-2024}%
  \BibitemOpen
  \bibfield  {author} {\bibinfo {author} {\bibfnamefont {L.~P.}\ \bibnamefont
  {Pryadko}}\ and\ \bibinfo {author} {\bibfnamefont {W.}~\bibnamefont {Zeng}},\
  }\href {https://github.com/QEC-pages/dist-m4ri} {\bibinfo {title} {{\tt
  dist\_m4ri} - distance of a classical or quantum {CSS} code}},\ \bibinfo
  {howpublished} {Published online} (\bibinfo {year} {2024})\BibitemShut
  {NoStop}%
\bibitem [{\citenamefont {Lin}\ \emph {et~al.}(2025)\citenamefont {Lin},
  \citenamefont {Liu}, \citenamefont {Lim},\ and\ \citenamefont
  {Pryadko}}]{Lin-Liu-Lim-Pryadko-github-2shot}%
  \BibitemOpen
  \bibfield  {author} {\bibinfo {author} {\bibfnamefont {H.-K.}\ \bibnamefont
  {Lin}}, \bibinfo {author} {\bibfnamefont {X.}~\bibnamefont {Liu}}, \bibinfo
  {author} {\bibfnamefont {P.~K.}\ \bibnamefont {Lim}},\ and\ \bibinfo {author}
  {\bibfnamefont {L.~P.}\ \bibnamefont {Pryadko}},\ }\href
  {https://github.com/QEC-pages/TwoShot} {\bibinfo {title} {{T}wo{S}hot
  repository}},\ \bibinfo {howpublished} {{G}it{H}hub repository} (\bibinfo
  {year} {2025})\BibitemShut {NoStop}%
\bibitem [{\citenamefont {Dumer}\ \emph {et~al.}(2015)\citenamefont {Dumer},
  \citenamefont {Kovalev},\ and\ \citenamefont
  {Pryadko}}]{Dumer-Kovalev-Pryadko-bnd-2015}%
  \BibitemOpen
  \bibfield  {author} {\bibinfo {author} {\bibfnamefont {I.}~\bibnamefont
  {Dumer}}, \bibinfo {author} {\bibfnamefont {A.~A.}\ \bibnamefont {Kovalev}},\
  and\ \bibinfo {author} {\bibfnamefont {L.~P.}\ \bibnamefont {Pryadko}},\
  }\bibfield  {title} {\bibinfo {title} {Thresholds for correcting errors,
  erasures, and faulty syndrome measurements in degenerate quantum codes},\
  }\href {https://doi.org/10.1103/PhysRevLett.115.050502} {\bibfield  {journal}
  {\bibinfo  {journal} {Phys. Rev. Lett.}\ }\textbf {\bibinfo {volume} {115}},\
  \bibinfo {pages} {050502} (\bibinfo {year} {2015})},\ \Eprint
  {https://arxiv.org/abs/1412.6172} {1412.6172} \BibitemShut {NoStop}%
\bibitem [{\citenamefont {Aasen}\ \emph
  {et~al.}(2025{\natexlab{a}})\citenamefont {Aasen}, \citenamefont {Haah},
  \citenamefont {Hastings},\ and\ \citenamefont
  {Wang}}]{Aasen-Haah-Hastings-Wang-2025}%
  \BibitemOpen
  \bibfield  {author} {\bibinfo {author} {\bibfnamefont {D.}~\bibnamefont
  {Aasen}}, \bibinfo {author} {\bibfnamefont {J.}~\bibnamefont {Haah}},
  \bibinfo {author} {\bibfnamefont {M.~B.}\ \bibnamefont {Hastings}},\ and\
  \bibinfo {author} {\bibfnamefont {Z.}~\bibnamefont {Wang}},\ }\bibfield
  {title} {\bibinfo {title} {Geometrically enhanced topological quantum
  codes},\ }\Eprint {https://arxiv.org/abs/2505.10403} {2505.10403}  (\bibinfo
  {year} {2025}{\natexlab{a}}),\ \bibinfo {note} {unpublished}\BibitemShut
  {NoStop}%
\bibitem [{\citenamefont {Aasen}\ \emph
  {et~al.}(2025{\natexlab{b}})\citenamefont {Aasen}, \citenamefont {Hastings},
  \citenamefont {Kliuchnikov}, \citenamefont {Bello-Rivas}, \citenamefont
  {Paetznick}, \citenamefont {Chao}, \citenamefont {Reichardt}, \citenamefont
  {Zanner}, \citenamefont {da~Silva}, \citenamefont {Wang},\ and\ \citenamefont
  {Svore}}]{Aasen-etal-Svore-2025}%
  \BibitemOpen
  \bibfield  {author} {\bibinfo {author} {\bibfnamefont {D.}~\bibnamefont
  {Aasen}}, \bibinfo {author} {\bibfnamefont {M.~B.}\ \bibnamefont {Hastings}},
  \bibinfo {author} {\bibfnamefont {V.}~\bibnamefont {Kliuchnikov}}, \bibinfo
  {author} {\bibfnamefont {J.~M.}\ \bibnamefont {Bello-Rivas}}, \bibinfo
  {author} {\bibfnamefont {A.}~\bibnamefont {Paetznick}}, \bibinfo {author}
  {\bibfnamefont {R.}~\bibnamefont {Chao}}, \bibinfo {author} {\bibfnamefont
  {B.~W.}\ \bibnamefont {Reichardt}}, \bibinfo {author} {\bibfnamefont
  {M.}~\bibnamefont {Zanner}}, \bibinfo {author} {\bibfnamefont {M.~P.}\
  \bibnamefont {da~Silva}}, \bibinfo {author} {\bibfnamefont {Z.}~\bibnamefont
  {Wang}},\ and\ \bibinfo {author} {\bibfnamefont {K.~M.}\ \bibnamefont
  {Svore}},\ }\bibfield  {title} {\bibinfo {title} {A topologically
  fault-tolerant quantum computer with four dimensional geometric codes},\
  }\Eprint {https://arxiv.org/abs/2506.15130} {2506.15130}  (\bibinfo {year}
  {2025}{\natexlab{b}}),\ \bibinfo {note} {unpublished}\BibitemShut {NoStop}%
\bibitem [{\citenamefont {Lin}\ \emph {et~al.}()\citenamefont {Lin},
  \citenamefont {Lim}, \citenamefont {Kovalev},\ and\ \citenamefont
  {Pryadko}}]{Lin-Lim-Kovalev-Pryadko-2025}%
  \BibitemOpen
  \bibfield  {author} {\bibinfo {author} {\bibfnamefont {H.-K.}\ \bibnamefont
  {Lin}}, \bibinfo {author} {\bibfnamefont {P.~K.}\ \bibnamefont {Lim}},
  \bibinfo {author} {\bibfnamefont {A.~A.}\ \bibnamefont {Kovalev}},\ and\
  \bibinfo {author} {\bibfnamefont {L.~P.}\ \bibnamefont {Pryadko}},\
  }\bibfield  {title} {\bibinfo {title} {Abelian multi-cycle codes for
  single-shot error correction},\ }\Eprint {https://arxiv.org/abs/2506.16910}
  {2506.16910} ,\ \bibinfo {note} {unpublished}\BibitemShut {NoStop}%
\bibitem [{\citenamefont {Yamasaki}\ and\ \citenamefont
  {Koashi}(2024)}]{Yamasaki-Koashi-2024}%
  \BibitemOpen
  \bibfield  {author} {\bibinfo {author} {\bibfnamefont {H.}~\bibnamefont
  {Yamasaki}}\ and\ \bibinfo {author} {\bibfnamefont {M.}~\bibnamefont
  {Koashi}},\ }\bibfield  {title} {\bibinfo {title} {Time-efficient
  constant-space-overhead fault-tolerant quantum computation},\ }\href
  {https://doi.org/10.1038/s41567-023-02325-8} {\bibfield  {journal} {\bibinfo
  {journal} {Nature Physics}\ }\textbf {\bibinfo {volume} {20}},\ \bibinfo
  {pages} {247} (\bibinfo {year} {2024})},\ \Eprint
  {https://arxiv.org/abs/2207.08826} {2207.08826} \BibitemShut {NoStop}%
\bibitem [{\citenamefont {Yoshida}\ \emph {et~al.}(2024)\citenamefont
  {Yoshida}, \citenamefont {Tamiya},\ and\ \citenamefont
  {Yamasaki}}]{Yoshida-Tamiya-Yamasaki-2024}%
  \BibitemOpen
  \bibfield  {author} {\bibinfo {author} {\bibfnamefont {S.}~\bibnamefont
  {Yoshida}}, \bibinfo {author} {\bibfnamefont {S.}~\bibnamefont {Tamiya}},\
  and\ \bibinfo {author} {\bibfnamefont {H.}~\bibnamefont {Yamasaki}},\
  }\bibfield  {title} {\bibinfo {title} {Concatenate codes, save qubits},\
  }\Eprint {https://arxiv.org/abs/2402.09606} {2402.09606}  (\bibinfo {year}
  {2024}),\ \bibinfo {note} {unpublished}\BibitemShut {NoStop}%
\end{thebibliography}%

\end{document}